\documentclass[reprint,amsmath,amssymb,aps,prl,superscriptaddress]{revtex4-1}

\usepackage{graphicx}
\usepackage{dcolumn}
\usepackage{bm}

\usepackage{xcolor}

\usepackage{subcaption}
\newcommand{\labelphantom}[1]{%
{\phantomsubcaption%
\label{#1}}%
}%

\begin{document}

\title{Characterization of the ELM-free Negative Triangularity Edge on DIII-D} 

\author{A.O. Nelson}
\email[Corresponding author: ]{a.o.nelson@columbia.edu}
\affiliation{Columbia University, New York City, New York, USA}

\author{L. Schmitz}
\affiliation{University of California - Los Angeles, Los Angeles, California, USA}

\author{T. Cote} 
\affiliation{General Atomics, San Diego, California, USA}

\author{J. F. Parisi}
\affiliation{Princeton Plasma Physics Laboratory, Princeton, New Jersey, USA}

\author{S. Stewart}
\affiliation{University of Wisconsin - Madison, Madison, Wisconsin, USA}

\author{C. Paz-Soldan}
\affiliation{Columbia University, New York City, New York, USA}

\author{K.E. Thome} 
\affiliation{General Atomics, San Diego, California, USA}

\author{M.E. Austin}
\affiliation{University of Texas – Austin, Austin, Texas, USA}

\author{F. Scotti}
\affiliation{Lawrence Livermore National Laboratory, Livermore, California, USA}

\author{J. L. Barr}
\affiliation{General Atomics, San Diego, California, USA}

\author{A. Hyatt}
\affiliation{General Atomics, San Diego, California, USA}

\author{N. Leuthold}
\affiliation{Columbia University, New York City, New York, USA}

\author{A. Marinoni}
\affiliation{University of California - San Diego, San Diego, California, USA}

\author{T. Neiser} 
\affiliation{General Atomics, San Diego, California, USA}

\author{T. Osborne}
\affiliation{General Atomics, San Diego, California, USA}

\author{N. Richner}
\affiliation{Oak Ridge Associated Universities, Oak Ridge, Tennessee, USA}


\author{A.S. Welander}
\affiliation{General Atomics, San Diego, California, USA}

\author{W.P. Wehner}
\affiliation{General Atomics, San Diego, California, USA}

\author{R. Wilcox}
\affiliation{Oak Ridge National Laboratory, Oak Ridge, Tennessee, USA}

\author{T. M. Wilks}
\affiliation{Massachusetts Institute of Technology, Cambridge, Massachusetts, USA}

\author{J. Yang}
\affiliation{Princeton Plasma Physics Laboratory, Princeton, New Jersey, USA}

\author{the DIII-D Team}

\begin{abstract}
Tokamak plasmas with strong negative triangularity (NT) shaping typically exhibit fundamentally different edge behavior than conventional L-mode or H-mode plasmas.
Over the entire DIII-D database, plasmas with sufficiently negative triangularity are found to be inherently free of edge localized modes (ELMs), even at injected powers well above the predicted L-H power threshold. 
A critical triangularly ($\delta_\mathrm{crit}\simeq-0.15$), consistent with inherently ELM-free operation is identified, beyond which access to the second stability region for infinite-$n$ ballooning modes closes on DIII-D. 
It is also possible to close access to this region, and thereby prevent an H-mode transition, at weaker average triangularities ($\delta\lesssim\delta_\mathrm{crit}$) provided that at least one of the two x-points is still sufficiently negative. 
Enhanced low field side magnetic fluctuations during ELM-free operation are consistent with additional turbulence limiting the NT edge gradient.
Despite the reduced upper limit on the pressure gradient imposed by ballooning stability, NT plasmas are able to support small pedestals and are typically characterized by an enhancement of edge pressure gradients beyond those found in traditional L-mode plasmas. 
Further, the pressure gradient inside of this small pedestal is unusually steep, allowing access to high core performance 
that is competitive with other ELM-free regimes previously achieved on DIII-D.
Since ELM-free operation in NT is linked directly to the magnetic geometry, NT fusion pilot plants are predicted to maintain advantageous edge conditions even in burning plasma regimes, potentially eliminating reactor core-integration issues caused by ELMs. 
\end{abstract}
\date{\today}

\maketitle


\section{Introduction}

The edge region of tokamak plasmas can be responsible for huge gains in plasma performance and/or triggering potentially dangerous plasma instabilities \cite{wagner_quarter-century_2007, Leonard2014}. Typically spanning the range $0.85\lesssim\psi_\mathrm{N}\leq1$, where the normalized poloidal flux $\psi_\mathrm{N}$ is a representative measure of the plasma radius, the tokamak edge also acts as the buffer between the hot core plasma and the cooler scrape-off-layer (SOL). Accordingly, optimization of the edge region is extremely important for scenario development, core-edge integration and power handling in tokamaks. However, this region is also notoriously difficult to model. Though intensive efforts to understand the tokamak edge have led to numerous important insights \cite{Snyder2009a, Urano2014, nelson_setting_2020}, the combination of steep temperature and density gradients, neutral sources, particle losses, strong magnetic and $E\times B$ shearing effects, diagnostic challenges and differences in SOL and core transport has to-date inhibited the creation of a fully-predictive and generalized model of the tokamak edge. As such, experimental characterizations of edge phenomena have great value in illuminating physics behaviors and informing scalings for next-step devices.


With that in mind, the goal of the present manuscript is to develop and document a comprehensive characterization of the edge region for DIII-D plasmas with strong negative triangularity (NT) shaping. 
In recent years, however, NT plasmas 
have emerged as an exciting candidate for pilot plant design by maintaining high core performance while alleviating constraints placed on the edge region \cite{marinoni_brief_2021}. Most notably, experiments on TCV and DIII-D have shown that, if the triangularity is \textit{sufficiently negative} (i.e. that triangularity $\delta$ is less than some critical value $\delta_\mathrm{crit}<0$), core confinement can be enhanced without triggering edge localized modes (ELMs) \cite{pochelon_energy_1999, Camenen2007, merle_pedestal_2017, Austin2019, marinoni_h-mode_2019, marinoni_diverted_2021, saarelma_ballooning_2021, nelson_prospects_2022, coda_enhanced_2022, nelson_robust_2023}. Since ELMs cannot be tolerated at the plasma currents forecasted for tokamak-based power plants \cite{Gunn2017, paz-soldan_plasma_2021, viezzer_prospects_2023}, this demonstrated compatibility between ELM-free operation and high core performance optimistically indicates that NT configurations could provide a safer ``power-handling first'' route towards fusion pilot plant (FPP) construction \cite{Medvedev2015, Kikuchi2019, frank_radiative_2022, schwartz_dee_2022, rutherford_manta_2024, guizzo_assessment_2024, miller_power_2024}. 

While NT is a potentially transformative scenario for the tokamak path to fusion energy, it is still under-studied compared to its positive triangularity (PT) counterpart. Building off a set of recent NT experiments executed on the DIII-D tokamak \cite{luxon_design_2002}, which are briefly described in section~\ref{sec:exp} and described in more depth in \cite{thome_overview_2024}, we help to close that gap here by documenting several important features of the NT edge as observed in strongly-shaped DIII-D plasmas. In particular, we discuss possible explanations for the inherently ELM-free nature of NT in section~\ref{sec:ELM}, demonstrating that the entire DIII-D dataset is consistent with the hypothesis that high-$n$ ballooning modes provide an upper limit for gradient growth in the NT edge that prevents destabilization of ELMs. We then quantify the underlying structure of the NT edge in section~\ref{sec:ped}, finding that substantial gradients can still be established in the NT edge while obeying the high-$n$ ballooning constraint. These observations suggest that the NT edge fundamentally differs from standard PT L-mode and H-mode plasmas. In light of this, we propose that these plasmas are more accurately described as having an ``ELM-free NT edge" rather than as being in ``L-mode.'' A brief summary of the characteristic turbulence signatures present in the strongly-shaped DIII-D NT edge are presented in section~\ref{sec:fluct}. Consequences of these findings, building off of the inherent nature of the NT edge to remain ELM-free even at high power, are presented as conclusions in  section~\ref{sec:conc} along with several potentially lucrative avenues of future study. 


\section{Experimental Methodology}
\label{sec:exp}

\begin{figure}
    \includegraphics[width=1\linewidth]{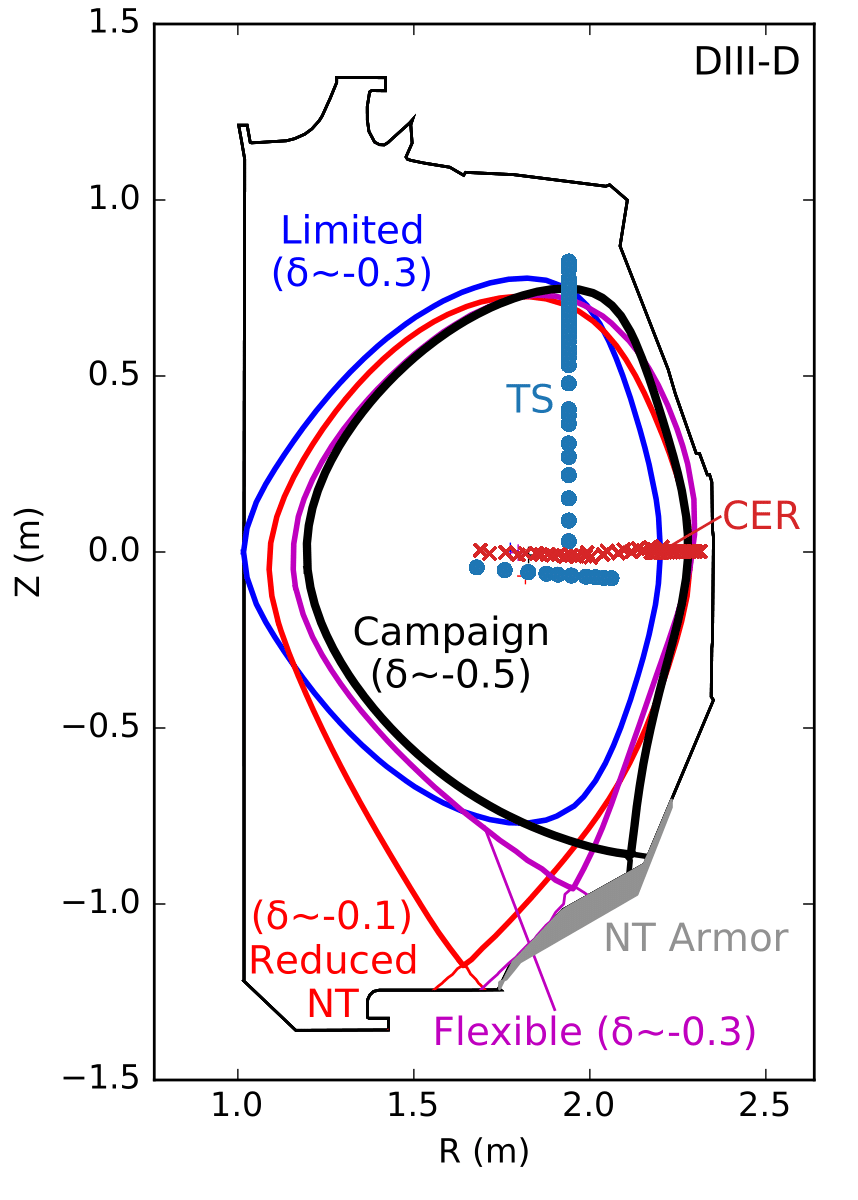}
    \caption{Various NT plasma shapes are presently achievable on DIII-D, providing a range of conditions for edge studies. Locations of the key Thomson scattering (TS) and charge-exchange recombination spectroscopy (CER) diagnostics used in this work are shown for the campaign configuration.}
    \label{fig:eqs}
\end{figure}

While not optimized with NT operation in mind, the flexibility of the DIII-D tokamak makes it an ideal location in which to study the effects of shaping on plasma operation. This paper reports on a broad array of DIII-D plasmas with triangularities $\delta<0$, as shown in figure~\ref{fig:eqs}. Here, and throughout this work, $\delta$ refers to the average of the top and bottom triangularities, which are denoted with $\delta_\mathrm{top}$ and $\delta_\mathrm{bottom}$, respectively. While the ``limited" (blue, \cite{Austin2019, marinoni_h-mode_2019}) and ``reduced NT" (red, \cite{marinoni_diverted_2021, saarelma_ballooning_2021}) shapes have been explored in the literature before, showing high performance ELM-free operation \cite{paz-soldan_plasma_2021}, a new set of water-cooled graphite tiles installed on the lower outboard side of DIII-D (grey, labeled ``NT armor") allows for the expansion of operation to two additional diverted configurations at strong $\delta<0$, as is discussed in more depth in \cite{nelson_vertical_2023, thome_overview_2024}. These are the ``flexible" shape (magenta), which has a somewhat scannable lower x-point to allow for variation in $\delta_\mathrm{lower}$ but is limited to heating powers of $P_\mathrm{aux}\lesssim4\,$MW
, and the ``campaign" shape (black), which has a fixed shape at $\delta\sim-0.5$ but can be maintained with the full heating power available on DIII-D. 

\begin{figure}
    \includegraphics[width=0.9\linewidth]{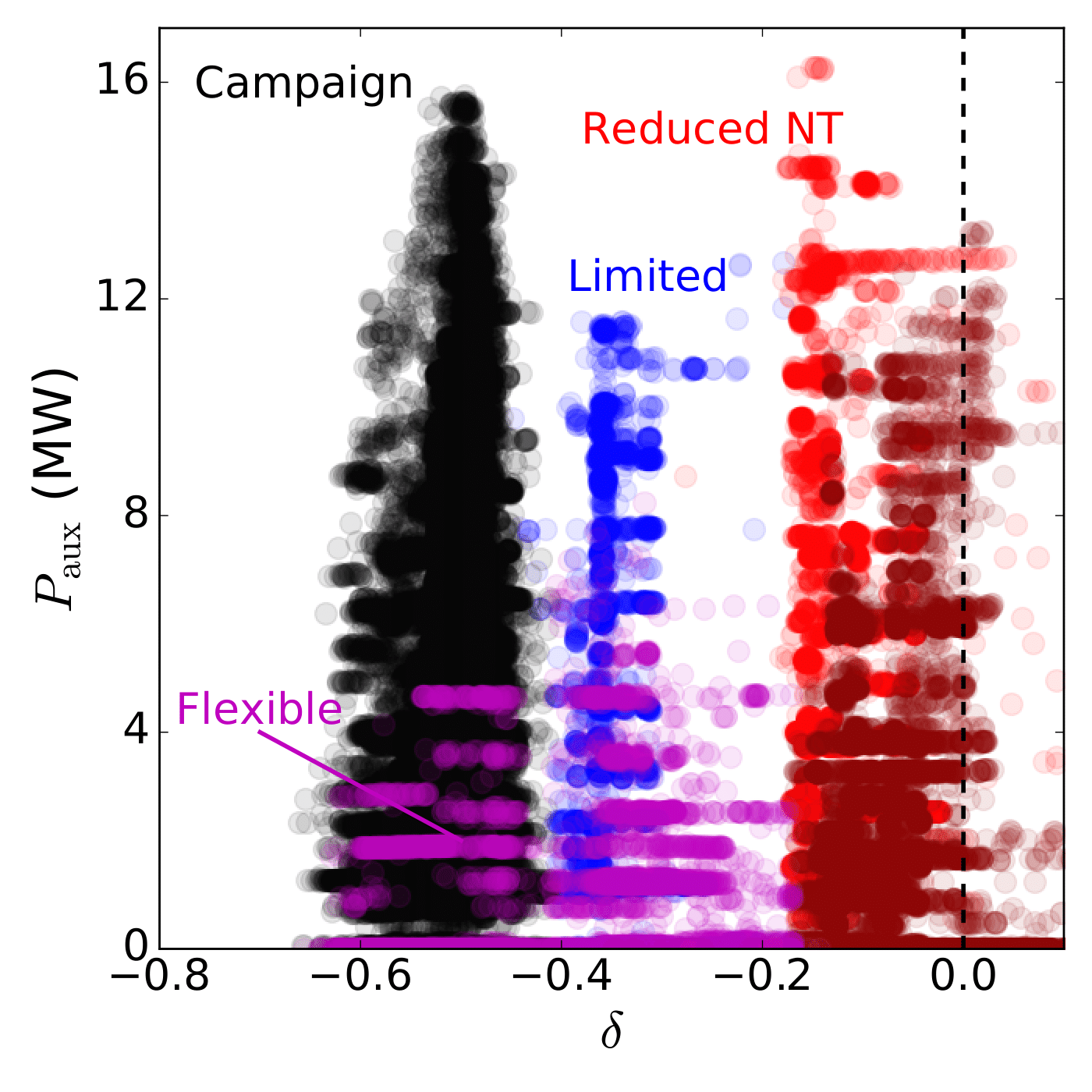}
    \caption{Total auxiliary heating power $P_\mathrm{aux}$ vs the average triangularity $\delta$ for all NT DIII-D discharges. Dark red points are also in the original ``reduced NT" shape, but not included in references \cite{marinoni_diverted_2021} and \cite{saarelma_ballooning_2021}.}
    \label{fig:p_vs_d}
\end{figure}

While results from all of these shapes will be discussed in this work, the campaign shape in particular fills an important role by expanding the DIII-D dataset to high-power, diverted discharges in strong NT, as depicted in figure~\ref{fig:p_vs_d}. DIII-D discharges in this shape covered a broad range of parameters, including plasma currents $I_\mathrm{p} = 0.5 – 1.2\,$MA, magnetic fields $|B_\mathrm{T}| = 1 – 2.1\,$T, edge safety-factors $q_\mathrm{95} = 2.4 – 7$, auxiliary heating powers $P_\mathrm{aux} = 0 – 16\,$MW, injected torques $T_\mathrm{inj}=-4-10\,$Nm, line-averaged densities $\langle n_\mathrm{e}\rangle = 0.1-1.4\times10^{20}$\,m$^{-3}$ and pulse lengths up to $\Delta t_\mathrm{pulse}\sim6.5$\,s \cite{thome_overview_2024, paz-soldan_simultaneous_2024}. Note that the average densities reported here, which utilize the line-averaged interferometer, may be up to $10-20\%$ larger than calibrations from Thomson scattering, though this does not impact any conclusions presented in this work \cite{thome_overview_2024}. The large variation present in this dataset allows for a robust assessment of the edge conditions present in strongly-shaped diverted NT plasmas, as discussed in sections~\ref{sec:ELM} and \ref{sec:ped} below. Notably, however, none of the shapes presented in figure~\ref{fig:eqs} feature a strongly-baffled or closed divertor shape, and most feature a short SOL connection length. This strongly impacts the SOL conditions present in these discharges (as discussed in depth in \cite{scotti_high_2024}) but is not responsible for the ELM-free nature of the achieved plasmas, as discussed in section~\ref{sec:ELM}.

Also central to this work are the high time- and spatial-resolution diagnostics present on the DIII-D device. The locations of two of these systems, the Thomson scattering (TS) \cite{Ponce-Marquez2010, Eldon2012} and charge-exchange recombination spectroscopy (CER) \cite{Chrystal2016} diagnostics, which measure the temperature and density of electrons and ions, respectively, are shown in figure~\ref{fig:eqs} for the campaign and flexible scenarios. Both of these systems have good coverage of the edge region. Additional SOL measurements provided by $D_\alpha$ filterscopes \cite{Colchin2003} 
are also featured heavily in this work to assess the (non-)ELMy nature of the NT edge. 

\section{ELM-free Operating Space}
\label{sec:ELM}

The most traditional high-performance tokamak plasma, the PT H-mode,  benefits from a boost in global confinement due to the formation of a region of steep edge pressure gradients called the pedestal. While large pedestals can lead to significant gains in normalized performance, they also almost inevitably destabilize coupled peeling-ballooning (PB) modes that trigger ELMs, periodically connecting the hot core plasma to the cooler edge region and depositing tremendous heat fluxes on the machine walls \cite{zohm_edge_1996, Leonard2014}. At the high plasmas currents and powers needed in an FPP, ELMs may be powerful enough to fatally damage machine components \cite{Gunn2017}, necessitating the establishment of ELM-free scenarios that are compatible with both burning core conditions and dissipative heat-exhaust solutions \cite{paz-soldan_plasma_2021, viezzer_prospects_2023}. Negative triangularity plasmas have emerged as a possible solution to this power-handling dilemma due to their inherently ELM-free nature, as described below. 

In the past decade, several different physics mechanisms have been proposed as being potentially responsible for the inherently ELM-free phenomenology exhibited by strongly-shaped NT plasmas. Broadly speaking, these mechanisms fall into two general categories: (1) instabilities that induce additional transport in the edge region, preventing the growth of pedestal gradients to levels that would otherwise trigger the onset of PB instabilities \cite{nelson_prospects_2022, nelson_robust_2023, marinoni_brief_2021, saarelma_ballooning_2021, merle_pedestal_2017, merlo_investigating_2015, yu_understanding_2023} or (2) mechanisms through which the threshold power for entering H-mode is significantly raised \cite{nishimura_computational_2020, singh_zonal_2022, singh_geometric_2023, kramer_full_2024}. In either case, plasmas that have \textit{sufficiently negative} triangularities have been experimentally demonstrated to robustly inhibit access to an ELMy H-mode state. While it is likely that many of the above effects contribute in potentially overlapping ways under various experimental conditions, the onset of the inherently ELM-free NT edge on DIII-D has been found to be consistent with the closure of the window to the second stability region of infinite-$n$ ballooning modes at a critical triangularity $\delta_\mathrm{crit}$ \cite{nelson_robust_2023}.

\subsection{Characterization of the DIII-D Data Set}

\begin{figure}
    \includegraphics[width=0.9\linewidth]{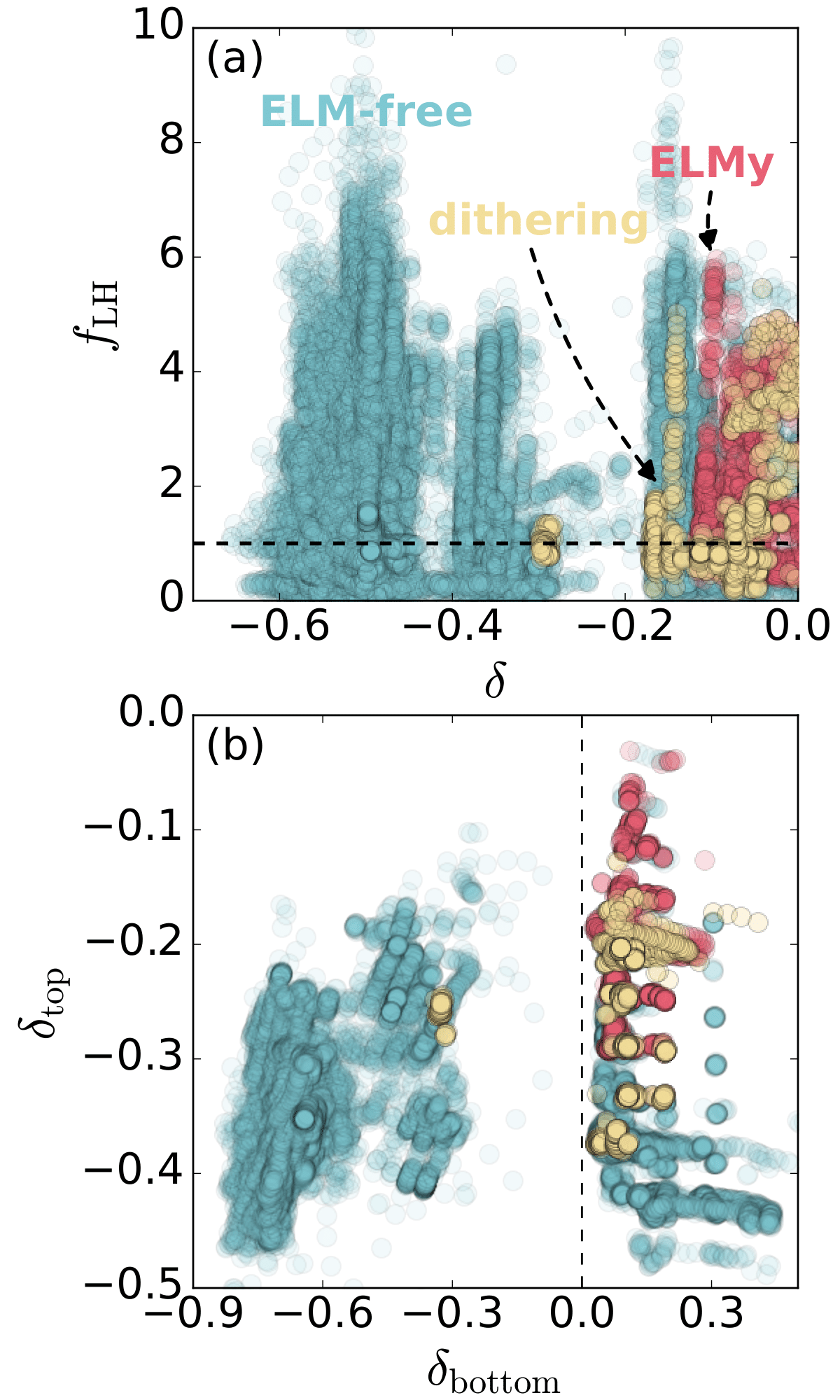}
    \labelphantom{fig:fLH-a}
    \labelphantom{fig:fLH-b}
    \caption{(a) Auxiliary power normalized to the expected L$\rightarrow$H transition power from equation~\ref{eq:Martin} vs. triangularity for every DIII-D discharge with $\delta<0$ (adapted from \cite{nelson_robust_2023}) All plasmas with $\delta<\delta_\mathrm{crit}\sim-0.15$ are inherently ELM-free, as indicated by the blue label. ELMy H-mode plasmas are colored red, and plasmas in the intermediate dithering regime are colored yellow. (b) The same set of discharges, sorted by the bottom and top triangularities ($\delta_\mathrm{bottom}$ and $\delta_\mathrm{top}$, respectively). 
    }
    \label{fig:fLH}
\end{figure}

\begin{figure}
    \includegraphics[width=1\linewidth]{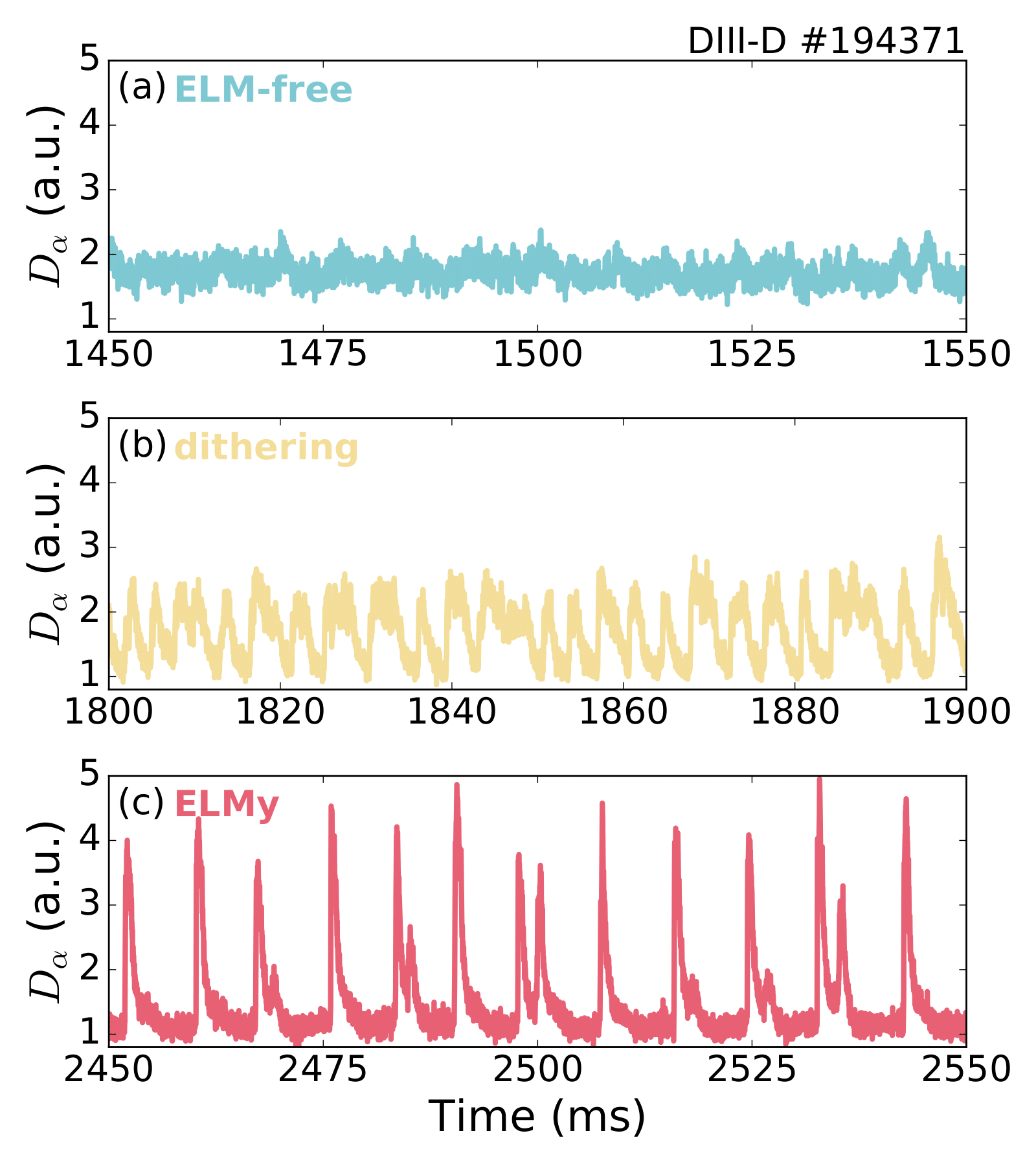}
    \labelphantom{fig:dalpha-a}
    \labelphantom{fig:dalpha-b}
    \labelphantom{fig:dalpha-c}
    \caption{$D_\mathrm{alpha}$ emission for (a) an ELM-free NT scenario, (b) a dithering scenario and (c) an NT H-mode with $0>\delta>\delta_\mathrm{crit}$. Colors are the same as in figure~\ref{fig:fLH}.}
    \label{fig:dalpha}
\end{figure}

\begin{figure}
    \includegraphics[width=1\linewidth]{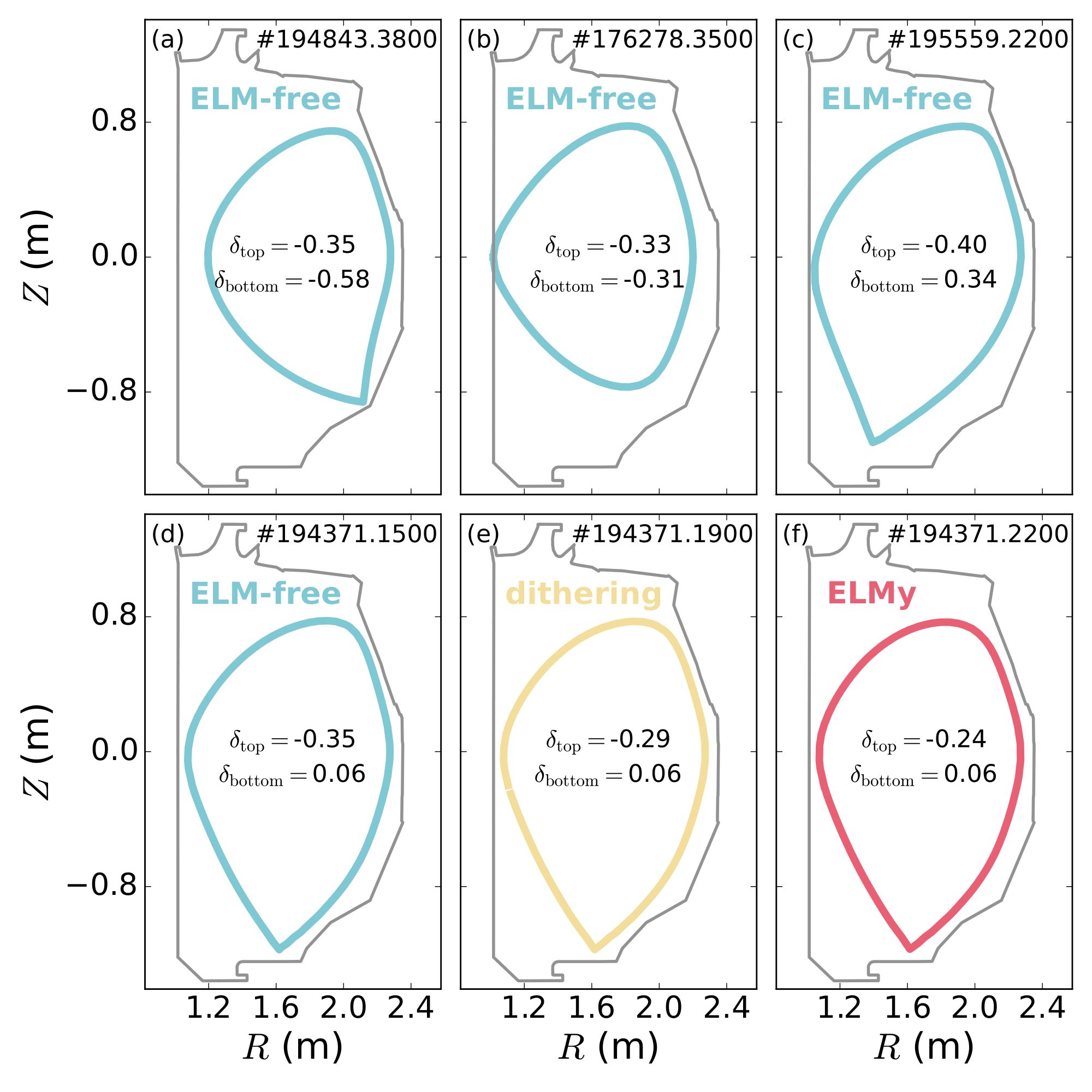}
    \labelphantom{fig:6eqs-a}
    \labelphantom{fig:6eqs-b}
    \labelphantom{fig:6eqs-c}
    \labelphantom{fig:6eqs-d}
    \labelphantom{fig:6eqs-e}
    \labelphantom{fig:6eqs-f}
    \caption{A series of DIII-D discharges, colored as in figure~\ref{fig:fLH} to show their ELMy nature. (a-c) An inherently ELM-free state is maintained over a wide range of shapes, as long as at least one of the two triangularities is sufficiently negative. (d-f) Near the critical triangularity, even small changes in the top triangularity can change trigger a loss of the ELM-free state.}
    \label{fig:6eqs}
\end{figure}

The concept of needing a sufficiently negative triangularity ($\delta<\delta_\mathrm{crit}$) in order to access the inherently ELM-free NT edge can be illustrated by considering the large database of NT discharges available from the DIII-D machine (figures~\ref{fig:eqs} and \ref{fig:p_vs_d}). For every discharge on DIII-D with at least one of the top and bottom triangularities ($\delta_\mathrm{top}$ and $\delta_\mathrm{bottom}$, respectively) less than zero, an assessment of the ELMy nature of the discharge is made every $20\,$ms by qualifying the $D_\mathrm{\alpha}$ signals provided by SOL filterscope measurements. This allows for a characterization of the ELM-free and ELMy conditions encountered over every plasma condition accessed in NT on the DIII-D tokamak. The results of this characterization are presented in figure~\ref{fig:fLH}, which colors each $20\,$ms time slice by its ELMy state: ELM-free times are marked in blue, ELMy times are marked in red, and ``dithery" times (described below) are marked in yellow. Representative $D_\mathrm{\alpha}$ traces for these three states are shown in figure~\ref{fig:dalpha}. For aid in visualization, cross-sections from a selection of these discharges are also included in figure~\ref{fig:6eqs}, which highlights the range of configurations accessible on DIII-D and their resulting ELM behavior. 

In figure~\ref{fig:fLH-a}, the DIII-D dataset is organized by averaged $\delta$ and the normalized loss power, which is defined as the fraction $f_\mathrm{LH}=P_\mathrm{loss}/P_\mathrm{LH08}$, where 
\begin{equation}
    P_\mathrm{loss} = P_\mathrm{aux} + P_\mathrm{Ohmic} - P_\mathrm{rad, core} - \frac{dW_\mathrm{MHD}}{dt}
    \label{eq:ploss}
\end{equation}
is a measure of the power crossing the separatrix and
\begin{equation}
    P_\mathrm{LH08} = 0.0488 \, \overline{n}^{0.717} B_\mathrm{t}^{0.803} S^{0.941}
    \label{eq:Martin}
\end{equation}
is the typical threshold power needed for H-mode access based on the scalings in reference~\cite{Martin2008}, which stem from a multi-machine database. Here the line-averaged plasma density $\overline{n}$ is given in [$10^{20}$ m$^{-3}$], $B_\mathrm{t}$ in [T], and the plasma surface area $S$ in [m$^2$]. Any plasma with $f_\mathrm{LH}\gtrsim1$ is expected to spontaneously transition into an ELMy H-mode state according to the conventional wisdom developed in PT configurations. Several key points are already realized by this presentation. Most importantly, we note that the critical triangularly needed to guarantee ELM-free operation is a finite negative value. In particular, there are plenty of examples on DIII-D (and other machines) of ELMy H-mode plasmas with $\delta<0$. However, no plasma on DIII-D that has $\delta<\delta_\mathrm{crit}\sim-0.15$ is able to achieve this transition: all DIII-D plasmas with a \textit{sufficiently negative} triangularity ($\delta < \delta_\mathrm{crit} < 0$) are inherently ELM-free, regardless of heating power or other plasma conditions. In fact several dedicated parametric variations designed to induced ELMs in this state were unsuccessful, highlighting the robustness of this phenomena. 

We also note that there exists a selection of points in figure~\ref{fig:fLH-a} that exist in an ELMy H-mode regime at $f_\mathrm{LH}<1$ and $-0.15<\delta<0$. Some of these time points are constituents of H-mode discharges in which the auxiliary power was dropped after H-mode was accessed, and H-mode operation thereafter sustained due to a known hysteresis of the L$\rightarrow$H power threshold \cite{kim_optimization_2022}. However, some of these discharges also spontaneously transition to H-mode at heating powers much less than the predicted value of $P_\mathrm{LH08}$. 
A more detailed characterization of this phenomena will be the subject of future work.

Importantly, it is not strictly necessary to have both $\delta_\mathrm{top}<\delta_\mathrm{crit}$ and $\delta_\mathrm{bottom}<\delta_\mathrm{crit}$ simultaneously in order to access the inherently ELM-free NT state. It is also possible to obtain inherently ELM-free operation if $\delta>\delta_\mathrm{crit}$, as long as either $\delta_\mathrm{top}$ or $\delta_\mathrm{bottom}$ is sufficiently negative to prevent access to the second stability region for infinite-$n$ ballooning modes. This can be seen directly in the DIII-D dataset, which is organized as a function of $\delta_\mathrm{top}$ and $\delta_\mathrm{bottom}$ in figure~\ref{fig:fLH-b}. Numerous ELM-free cases exist within the DIII-D dataset with $\delta_\mathrm{bottom} > 0$. Typically these discharges also exhibit $\delta_\mathrm{top} \ll 0$, such that the average triangularity $\delta$ remains strongly negative, though ELM-free cases with $\delta\sim0$ also exist within the dataset. One such example is depicted in figure~\ref{fig:6eqs-c}, which shows the cross-section for an inherently ELM-free NT plasma with $\delta=-0.06$, which is well under the ``global" $\delta_\mathrm{crit}$ reported in \cite{nelson_robust_2023}. In this case, the plasma is prevented from accessing H-mode via the strongly negative top triangularity only ($\delta_\mathrm{top}=-0.4$), consistent with an infinite-$n$ ballooning mode theory that requires only that one of the two x-points be sufficiently negative to prevent H-mode access. ELM-free plasmas with roughly opposite upper and lower triangularities have also been established on TCV, where single null plasmas with $\delta\ll0$ on the active x-point and $\delta>0$ on the inactive corner were found to exhibit enhanced performance \cite{sauter_negative_2023}. 

It is important to note here that it is theoretically and experimentally possible to reach H-mode in plasmas that live in the first stability region for infinite-$n$ ballooning modes. This phenomena is more common in low aspect ratio devices or in plasmas with collisionalities high enough to suppress the bootstrap current and has not (yet) been observed in DIII-D NT plasmas despite an extensive search. In such cases, NT plasmas may still remain ELM-free due to additional gradient-limiting mechanisms similar to those already observed on DIII-D, as discussed below.

Also present in the database plots presented in figure~\ref{fig:fLH} are a class of discharges that are characterized by a ``dithering" or ``limit-cycle-oscillation-like" behavior, which is typically present near the critical triangularity needed for second stability closure \cite{yu_understanding_2023, nelson_robust_2023}. In this state, the plasma occasionally exhibits enhanced pedestal gradients reminiscent of the limit cycle oscillations (LCOs) typically observed before an H-mode transition in PT \cite{staebler_h-mode_2015}. However, in none of these cases is the enhanced gradient ever sustained for more than $\sim20\,$ms, such that the time-averaged profiles are only characterized by a slight improvement in the pedestal gradient below what would be needed to trigger a peeling-ballooning instability. Additional work to further characterize this plasma state is needed. 

We note here also that the exact value of $\delta_\mathrm{crit}$ is not necessarily identical for each discharge on DIII-D, and is certainly not identical for NT discharges on different machines \cite{nelson_robust_2023}. Additionally, at triangularities near $\delta_\mathrm{crit}$, it is likely that various additional physics mechanisms (potentially including the $\nabla B$ drift direction, electron temperature gradient modes \cite{merlo_investigating_2015, marinoni_brief_2021}, kinetic ballooning modes \cite{merle_pedestal_2017}, micro-tearing modes \cite{nelson_time-dependent_2021} and/or zonal flow screening \cite{singh_zonal_2022, singh_geometric_2023}) will play a significant role in determining whether or not the plasma enters an ELMy H-mode state. This is consistent with data from AUG, which shows a strong dependence of the $\nabla B$ drift direction on H-mode access in configurations with $\delta\sim\delta_\mathrm{crit}$ \cite{happel_overview_2022, nelson_robust_2023}. It is important to note that, regardless of the particular plasma conditions, it is always possible on DIII-D to leave the dithering regime and access a robust ELM-free state by decreasing either $\delta_\mathrm{top}$ or $\delta_\mathrm{bottom}$ below a critical value. To avoid any complications that may occur as a result of these macroscopic oscillations, an NT FPP should simply be designed for a triangularity that is sufficiently negative not only to avoid ELMy H-mode operation, but also to robustly avoid any dithering behavior.

\subsection{Infinite-$n$ Ballooning Stability as an Upper Limit for the NT Edge Gradient}
\label{sec:ball}

\begin{figure}
    \includegraphics[width=1\linewidth]{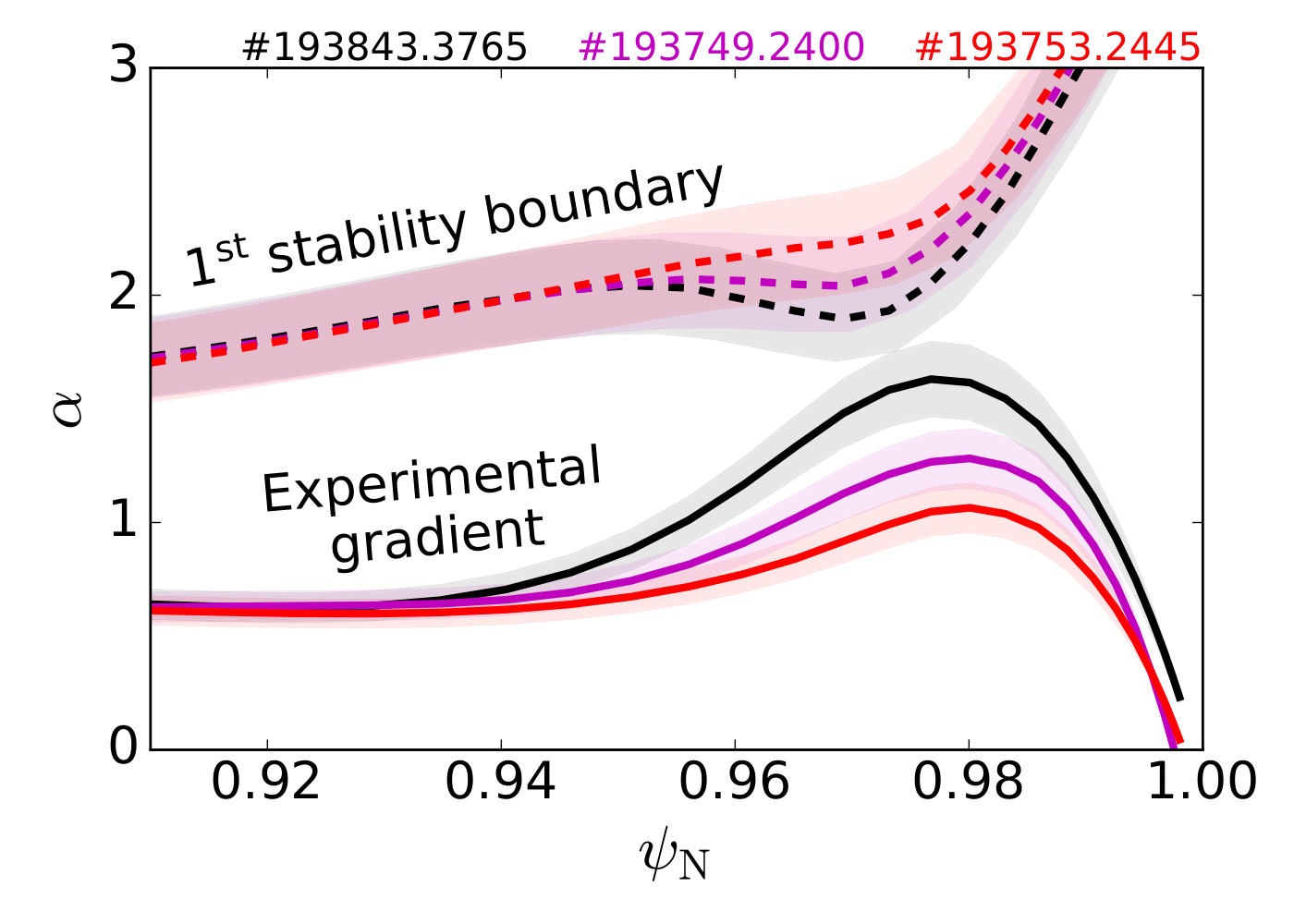}
    \caption{Infinite-$n$ ballooning stability calculations for three ELM-free NT plasmas on DIII-D, demonstrating that the experimental gradients observed in the edge are not always ballooning-limited. All three of these plasmas are in the campaign shape with $\delta=-0.5$.}
    \label{fig:ballex}
\end{figure}

The sharp cutoff in access to the inherently ELM-free observed above is largely consistent with the closure of the window to the second stability region of infinite-$n$ ballooning modes at a critical triangularity $\delta_\mathrm{crit}$ \cite{merle_pedestal_2017, saarelma_ballooning_2021, marinoni_brief_2021, nelson_prospects_2022, nelson_robust_2023, zhang_analytical_2024}. As described in-depth in previous work (especially in references \cite{nelson_prospects_2022, nelson_robust_2023, zhang_analytical_2024}), infinite-$n$ ballooning stability depends explicitly on the magnetic shear in the plasma edge, which in turn is a function of the internal plasma profiles, the plasma collisionality and the discharge geometry. Notably, as shown in figure 14 of reference \cite{nelson_prospects_2022}, for each set of machine conditions there exists some $\delta_\mathrm{crit}$ for both $\delta_\mathrm{top}$ or $\delta_\mathrm{bottom}$ past which access to the second stability region in closed, preventing growth of the pedestal pressure gradient to levels typically characteristic of an ELMy H-mode. Experimentally, this phenomenon manifests itself by establishing a sharp cutoff for ELMy H-mode access as a function of $\delta$, as observed in figure~\ref{fig:fLH}. Ballooning stability analysis presented in \cite{saarelma_ballooning_2021} and \cite{nelson_robust_2023} shows that the experimentally observed $\delta_\mathrm{crit}$ on DIII-D is consistent with value of $\delta$ needed to close access to the second stability window, supporting the idea that the infinite-$n$ ballooning mode is the key physics phenomena responsible for ensuring access to an inherently ELM-free state in strong NT. 

However, ELM-free NT plasmas on the DIII-D tokamak do not always exhibit pressure gradients that reach marginal ballooning stability, which can be directly calculated using a magneto-hydrodynamic (MHD) stability code such as BALOO \cite{Snyder2002}. To illustrate this point, radial profiles of the normalized experimental pressure gradient ($\alpha$) and the first stability boundary for infinite-$n$ ballooning modes are presented in figure~\ref{fig:ballex} for a series of DIII-D discharges in the strongly-shaped campaign shape (figure~\ref{fig:6eqs-a}). Here $\alpha$ is defined as
\begin{equation}
    \label{eq:alpha}
    \alpha = \frac{\mu_\mathrm{0}}{2\pi^2}\frac{\partial V}{\partial\psi}\bigg(\frac{V}{2\pi^2R_\mathrm{0}}\bigg)^{1/2}\frac{dp}{d\psi},
\end{equation}
where $V$ is the volume enclosed by each flux surface, $\psi$ the poloidal flux, $p$ the plasma pressure and $R_\mathrm{0}$ the plasma major radius. In all three cases, the pressure profile exists entirely within the first stability region for infinite-$n$ ballooning modes and does NOT achieve the requisite pressure gradients needed to trigger peeling-ballooning (PB) modes. However, while all of these plasmas have $f_\mathrm{LH}>1$, meaning that they would be expected to transition to an H-mode state in PT, only one of the three examples (\#193843 - black) supports a pressure gradient that is marginal to (and thus potentially limited by) the first stability boundary for infinite-$n$ ballooning modes. As such, we refer to the first stability boundary for infinite-$n$ ballooning only as an upper limit for the edge pressure gradient in NT plasmas, robustly preventing the plasmas from accessing PB-unstable conditions but not necessarily responsible for the absolute pressure gradient observed in each individual plasma. 

\begin{figure}
    \includegraphics[width=1\linewidth]{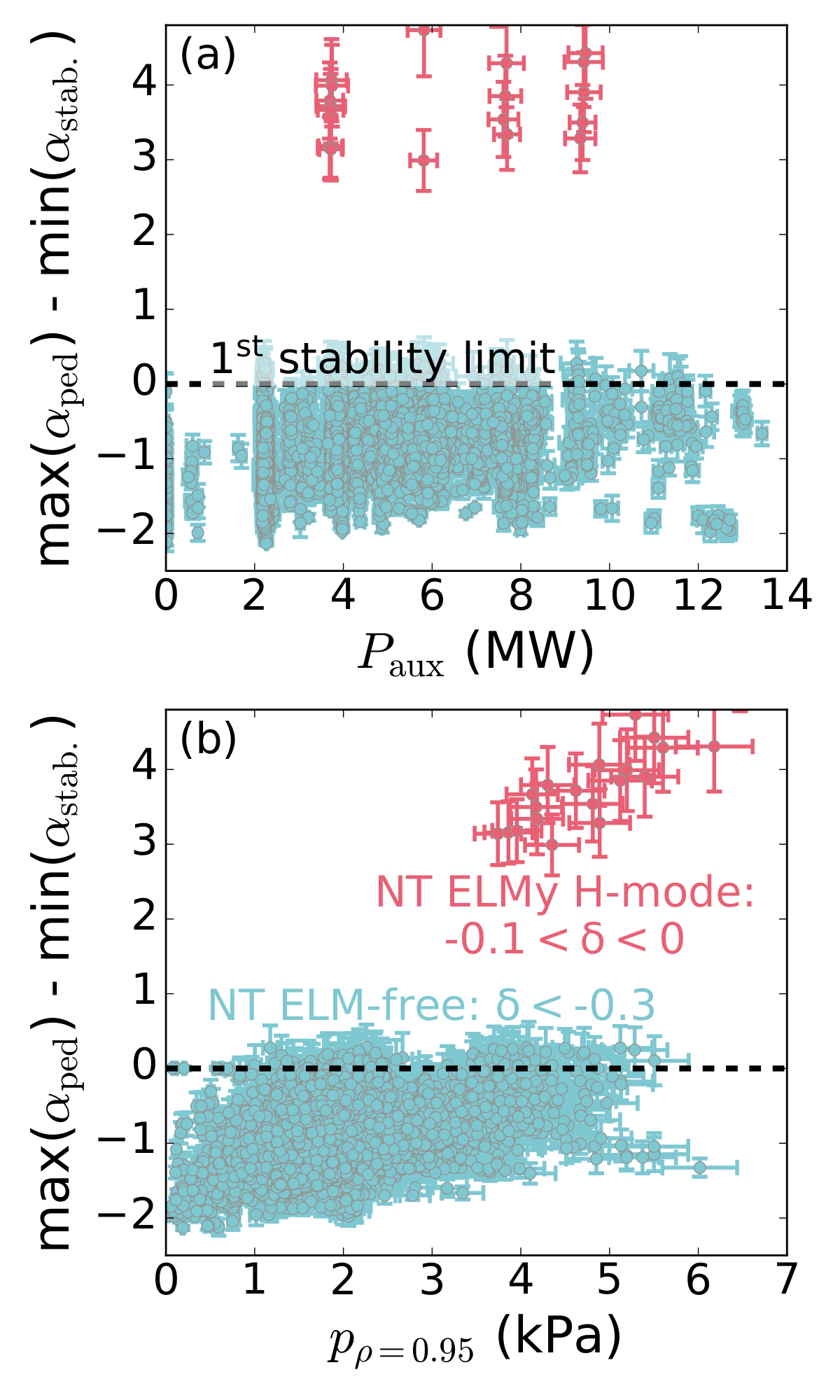}
    \labelphantom{fig:balldb-a}
    \labelphantom{fig:balldb-b}
    \caption{Proximity of the experimental pressure profile to the first ballooning stability boundary ($\mathrm{max}(\alpha_\mathrm{ped})-\mathrm{min}(\alpha_\mathrm{ballooning})$) as a function of (a) the auxiliary power and (b) the edge pressure, which is adapted from \cite{nelson_robust_2023}. Inherently ELM-free discharges are colored blue and a selection of ELMy H-modes obtained a weaker delta are included in red for comparison.}
    \label{fig:balldb}
\end{figure}

The consistency of the infinite-$n$ ballooning stability boundary as an upper limit for the pressure gradient in discharges with $\delta<\delta_\mathrm{crit}$ can be checked for the entire NT database on DIII-D by creating kinetic equilibria for each discharge using the CAKE code \cite{xing_cake_2021} and modeling their respective ballooning stability with BALOO. This is demonstrated in figure~\ref{fig:balldb}, which shows the distance between the maximum equilibrium $\alpha$ ($\mathrm{max}(\alpha_\mathrm{ped})$) and the minimum $\alpha$ required for ballooning mode destabilization ($\mathrm{min}(\alpha_\mathrm{ballooning})$) as a function of the auxiliary power ($P_\mathrm{aux}$) and the edge pressure ($p_\mathrm{\rho=0.95}$, where $\rho$ is the normalized radius). While no ELM-free discharges support larger gradients than allowed by ideal ballooning stability, there is a wide range in both the achieved edge pressure and the ballooning stability metric across the DIII-D NT database. Further, variation in the ballooning stability metric $\mathrm{max}(\alpha_\mathrm{ped})-\mathrm{min}(\alpha_\mathrm{ballooning})$ is not strongly correlated with the either the heating power or the edge pressure. This observed variation is consistent with additional physics mechanisms playing a role in setting the NT edge structure while the infinite-$n$ ballooning mode sets an absolute upper limit on the pressure gradient in the NT edge. To illustrate this point, a selection of strong ELMy H-mode discharges at weak $\delta<0$ (from \cite{saarelma_ballooning_2021, marinoni_diverted_2021, nelson_robust_2023}) are also shown in figure~\ref{fig:balldb} in red. In all of these cases, which include examples at lower $P_\mathrm{aux}$ and $p_\mathrm{\rho=0.95}$ that obtained in the NT database, the plasma is able to access the 2$^{nd}$ stability region and can therefore achieve normalized gradients well above the 1$^{st}$ stability limit. 


\subsection{Correlation Between the NT ELM-free Edge and a High-Performance Core}
\label{subsec:core}

\begin{figure}
    \includegraphics[width=1\linewidth]{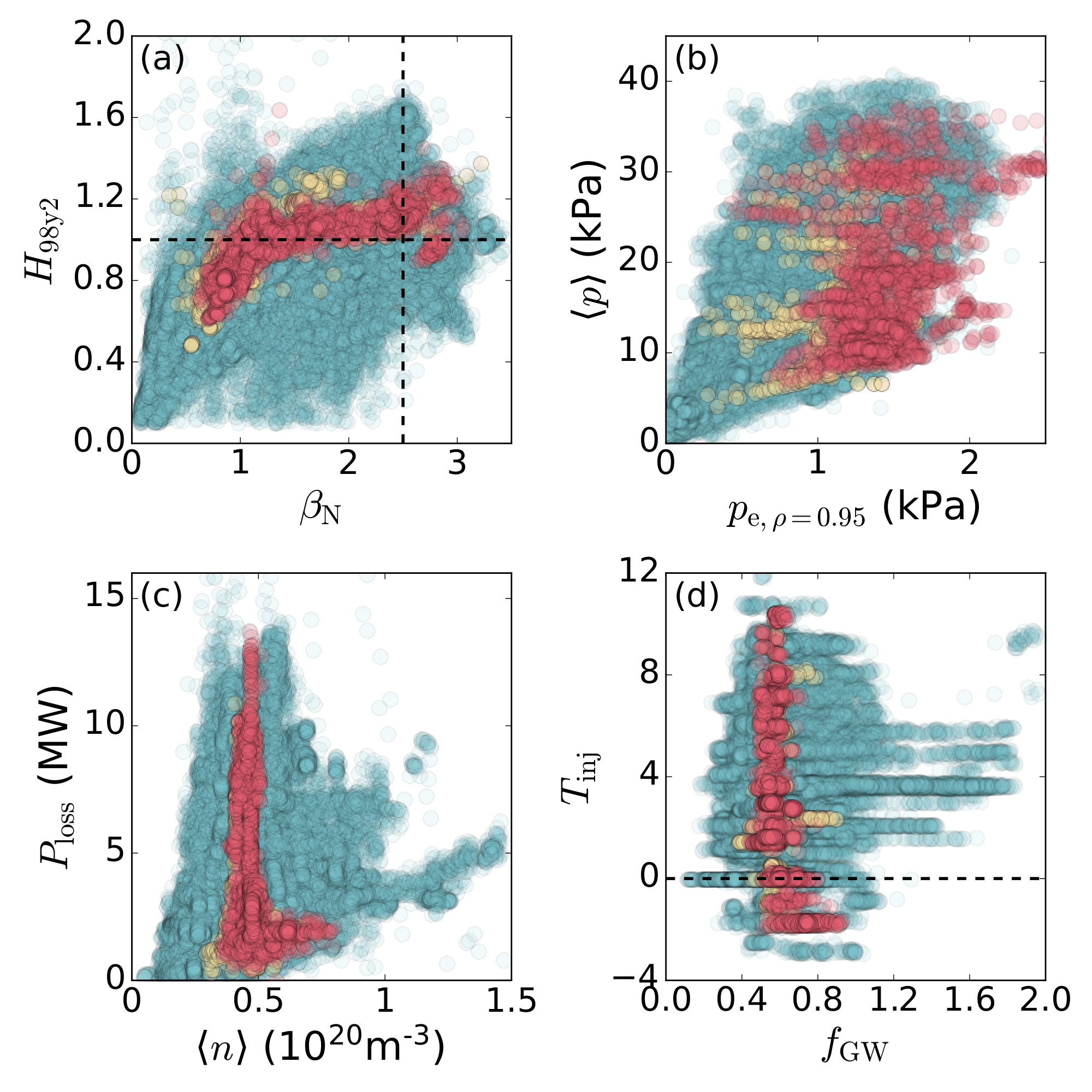}
    \labelphantom{fig:pow-a}
    \labelphantom{fig:pow-b}
    \labelphantom{fig:pow-c}
    \labelphantom{fig:pow-d}
    \caption{(a) $H_\mathrm{98y2}$ vs. $\beta_\mathrm{N}$, (b) volume-averaged pressure vs. edge pressure, (c) $P_\mathrm{loss}$ vs. volume-averaged density (adapted from \cite{nelson_robust_2023}) and (d) injected torque vs. $f_\mathrm{GW}$ for the entire DIII-D dataset. Colors represent the ELMy state: blue is ELM-free, yellow is dithery and red is ELMy, as in figure~\ref{fig:fLH}.}
    \label{fig:pow}
\end{figure}

One of the key distinguishing features of NT plasmas is that their robust ELM-free nature is compatible with numerous core solutions, including many that push performance records obtained in any ELM-free regime on DIII-D. To demonstrate this, we show several measures of reactor compatibility measured over all DIII-D plasmas with $\delta<0$ in figure~\ref{fig:pow}. Measures of the normalized conferment $H_\mathrm{98y2}$ (defined in \cite{ITERPhysicsBasisEditors1999}) and normalized pressure $\beta_\mathrm{N}$ are provided in figure~\ref{fig:pow-a}, showing that the inherently ELM-free NT edge is compatible with simultaneous access to $H_\mathrm{98y2}>1$ and $\beta_\mathrm{N}>2.5$, which should be sufficient for inductive FPP operation \cite{Austin2019, paz-soldan_simultaneous_2024, rutherford_manta_2024}. Notably, it is also evident from figure~\ref{fig:pow-a} that accessing inherently ELM-free conditions via NT does not necessarily result in a reduction in normalized performance. There are numerous ELM-free NT plasmas on DIII-D with higher $H_\mathrm{98y2}$ and $\beta_\mathrm{N}$ than any of the ELMy H-mode cases achieved at $\delta_\mathrm{crit}<\delta<0$, which are highlighted in red throughout figure~\ref{fig:pow}. 

A similar story can be seen when looking at absolute measures of the plasma pressure in figure~\ref{fig:pow-b}. While none of the ELM-free plasmas at strong $\delta$ can achieve the same pedestal pressures as the strongest ELMy H-mode cases, the inherently ELM-free plasmas are able to match or even exceed the volume-averaged pressures $\langle p \rangle$ achieved in NT H-modes at $\delta_\mathrm{crit}<\delta<0$. In a similar manner, ELM-free NT plasmas are able to compete with the established QH-mode and RMP ELM-suppressed H-mode regimes in terms of $\langle p \rangle$, as detailed in reference \cite{nelson_robust_2023}. One of the strengths of the NT regime is that it shifts some of the burden on plasma performance from the edge region to the core region, allowing for access to reactor-relevant conditions with relatively modest edge pressures, which may be beneficial for core-edge integration solutions in an FPP. 

Figure~\ref{fig:pow-c} provides a view of the total power crossing the separatrix ($P_\mathrm{loss}$) vs the plasma density. In PT plasmas, comparison of these variables can illuminate the density dependence of the L$\rightarrow$H transition power. A similar dependence is seen in figure~\ref{fig:pow-c}, where the $P_\mathrm{loss}$ required to enter into H-mode has a clear minimum around $\langle n \rangle=0.45\times10^{20}\,$m$^{-3}$. The inherently ELM-free discharges cover a space much broader than that accessed in the ELMy H-mode regime, demonstrating the robustness of the ELM-inhibition phenomenon. 

A comparison of the injected torque ($T_\mathrm{inj}$) and the Greenwald density fraction ($f_\mathrm{GW}$) is presented in figure~\ref{fig:pow-d}. By exploiting the increased density threshold at high power \cite{giacomin_first-principles_2022, hong_characterization_2023}, it is possible to achieve Greenwald fractions of nearly $f_\mathrm{GW}\sim2$ in the inherently ELM-free NT state \cite{hong_characterization_2023}. This emphasizes the flexibility of NT configurations to sustain heavy modifications of the core plasma without loosing their favorable ELM-free conditions and potentially suggest that a highly radiative, high density core solution may be achievable on an NT FPP \cite{rutherford_manta_2024, miller_power_2024}.


\section{Characterization of the NT Pedestal}
\label{sec:ped}

Since the height of the NT pedestal is not uniquely determined by the infinite-$n$ ballooning limit, it is prudent to consider other factors that may impact the structure of profiles in the NT edge. In this section, we document typical experimental characteristics of the NT edge region exhibited across the existing DIII-D NT database. In general, we observe that the NT pedestal on DIII-D can be quite narrow, but that the gradients just inside of the pedestal do not typically differ too substantially from the pedestal gradients themselves. This leads to an interesting behavior where large edge pressures (eg. $p(\psi_\mathrm{N}=0.9)$) can be sustained even with relatively modest pedestal heights. Because of this, the edge pressure in NT discharges can contribute significantly to increases in core performance.  

\subsection{Pedestal Structure}
\label{subsec:power}

While changes in core turbulent transport can contribute to the increased confinement exhibited by NT regimes \cite{merlo_investigating_2015, Camenen2007, Austin2019, marinoni_diverted_2021}, a large portion of the confinement improvement observed in NT plasmas can also be attributed to an enhancement of the edge gradient \cite{Sauter2014}. This is a result of the non-stiff nature of the NT edge region, which allows for the formation of enhanced gradients and pedestals in the NT edge \cite{Sauter2014} that are significantly steeper than those expected in traditional L-mode-like plasmas. In this respect, the edge region of NT plasmas is qualitatively different than both H-mode and L-mode scenarios in PT, necessitating the development of new predictive models and scalings laws that will be the subject of future work.

\begin{figure}
    \includegraphics[width=1\linewidth]{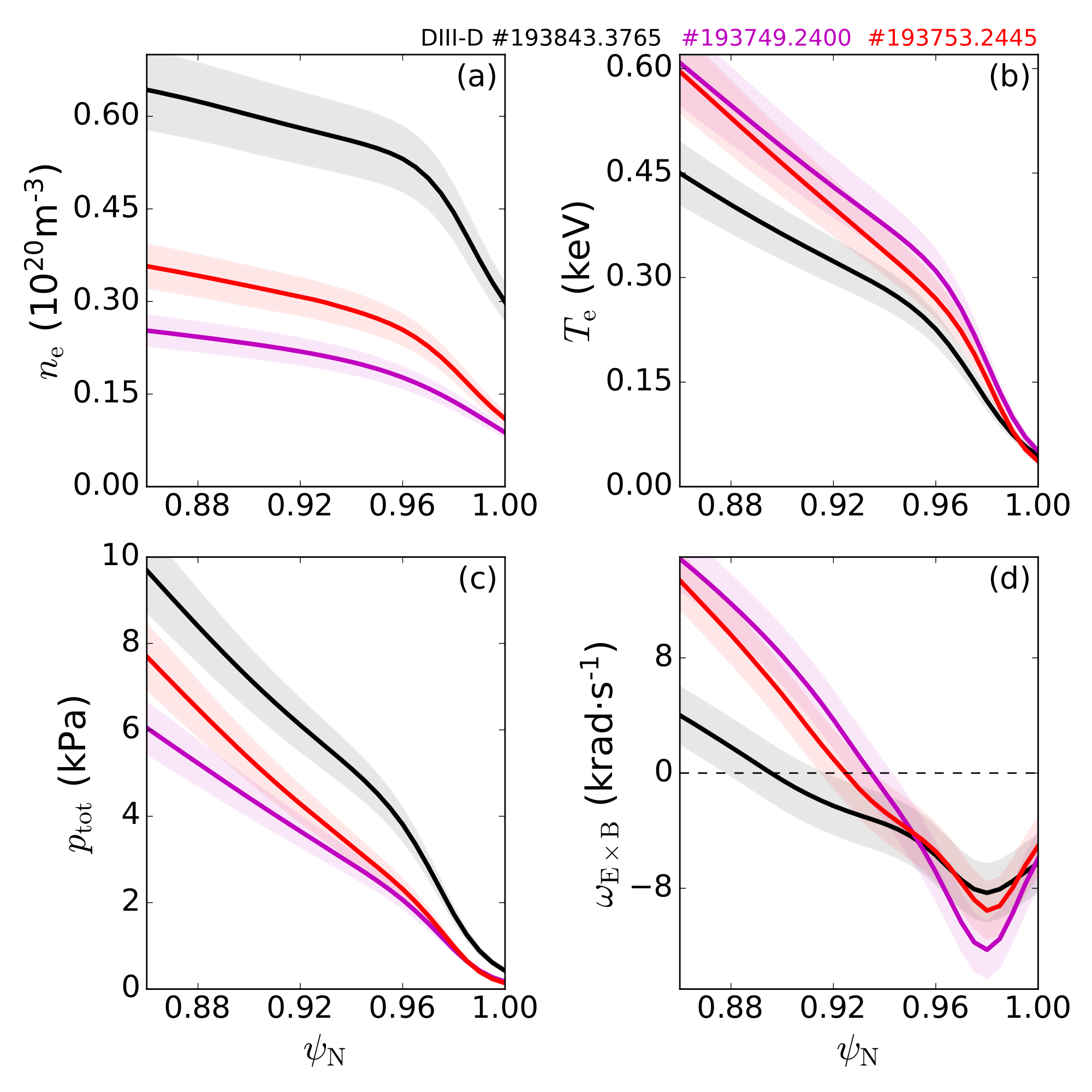}
    \labelphantom{fig:profs-a}
    \labelphantom{fig:profs-b}
    \labelphantom{fig:profs-c}
    \labelphantom{fig:profs-d}
    \caption{Edge (a) electron density, (b) electron temperature, (c) total pressure and (d) $E\times B$ drift velocity profiles in the edge region of the three NT plasmas from figure~\ref{fig:ballex}.}
    \label{fig:profs}
\end{figure}

An illustration of typical edge profiles in DIII-D NT plasmas is provided in figure~\ref{fig:profs}, using the same equilibria analyzed during the preceding discussion regarding figure~\ref{fig:ballex}. The black case, which is marginal to ballooning first stability, is from a high density, high power regime with a line-averaged density of $\langle n\rangle=8.2\times10^{20}\,$m$^{-3}$, $P_\mathrm{aux}=8.2\,$MW and plasma current $I_\mathrm{p}=1\,$MA. The purple and red cases are both from plasmas with $P_\mathrm{aux}=4\,$MW but have $\langle n\rangle=3.7\times10^{20}\,$m$^{-3}$ and $\langle n\rangle=4.7\times10^{20}\,$m$^{-3}$, respectively. The purple case is also at $I_\mathrm{p}=1\,$MA, whereas the red case has $I_\mathrm{p}=0.8\,$MA. Notably, all of these plasmas, despite the variation in their conditions, demonstrate a significantly enhanced electron temperature ($T_\mathrm{e}$) gradient in the edge region. A smaller gradient enhancement can be observed in the electron density ($n_\mathrm{e}$) profile, but the density profile remains mostly flat out to the separatrix in all but the highest density case. This is characteristic of typical NT plasmas on DIII-D, which often develop a relatively prominent $T_\mathrm{e}$ pedestal while the $n_\mathrm{e}$ profile remains flatter throughout the edge region. Notably, this is qualitatively similar to the I-mode edge, which is also characterized by the development of a thermal transport barrier while particle transport remains L-mode-like \cite{Whyte2010, hubbard_multi-device_2016}. I-mode plasmas, however, are also characterized by the presence of a weakly coherent mode, by an access condition requiring unfavorable $B\times\nabla B$ drift direction, and by an eventual threshold power $P_\mathrm{I H}$ above which a transition to H-mode is achieved. None of these specific I-mode characteristics are observed in NT plasmas on DIII-D, so we do not classify the NT edge as being an I-mode edge. However, the regimes do share many similarities in terms of pedestal structure. Further, as seen in figure~\ref{fig:profs}, the gradient inside of the pedestal in NT plasmas can be almost as strong as the pedestal itself, which is not typically true for I-mode plasmas. 

In figure~\ref{fig:profs-d}, we also show the $E\times B$ drift velocity in the edge region. Unlike in PT L-mode plasmas, the ELM-free NT edge can support the growth of a radial electric field ($E_\mathrm{r}$) well near the separatrix that can contribute to significant levels of $E\times B$ shear. The growth of the $E_\mathrm{r}$ well in ELM-free NT plasmas is correlated with strong beam heating. 

\begin{figure}
    \includegraphics[width=1\linewidth]{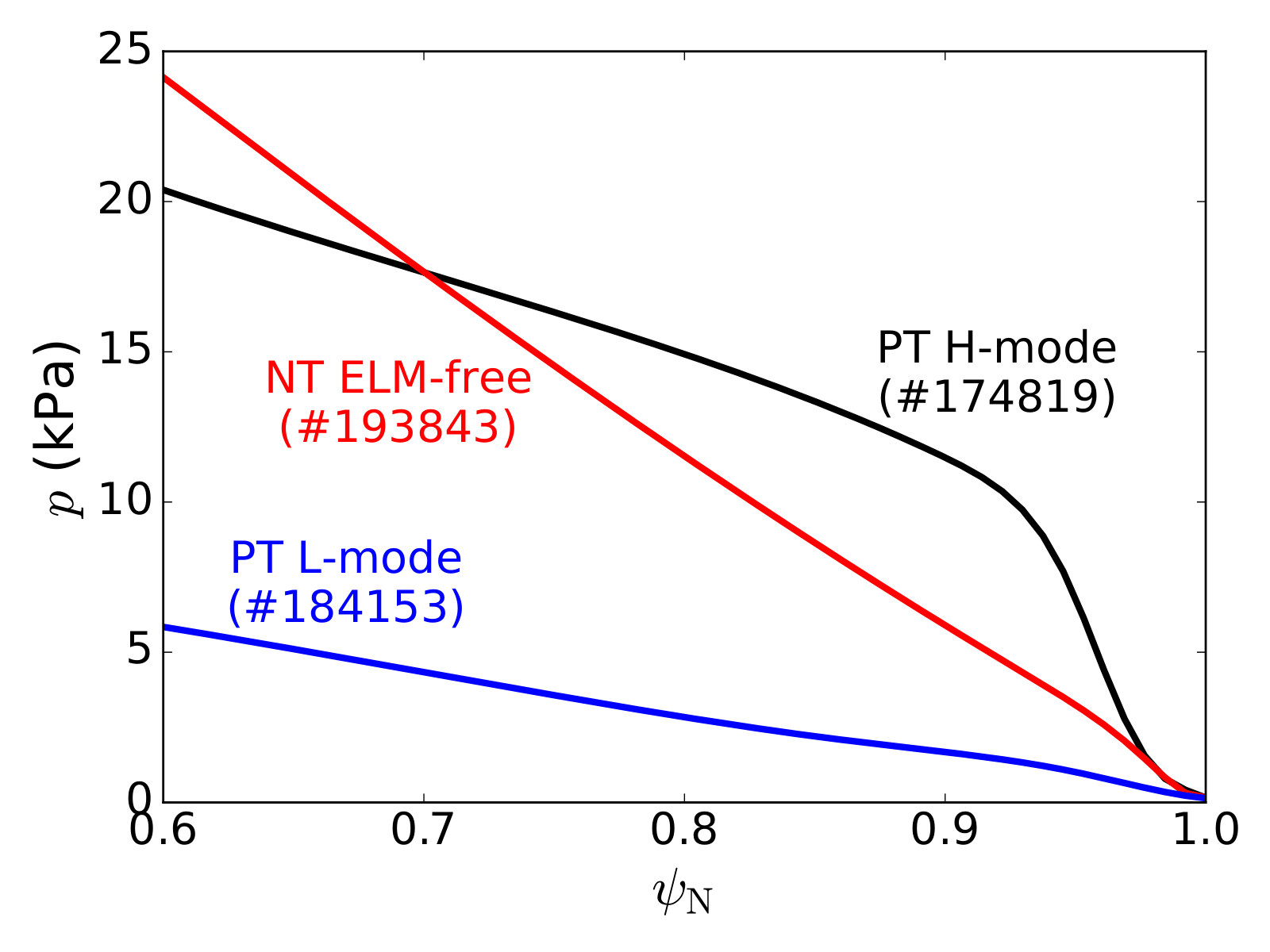}
    \caption{Pressure profiles for example matched PT H-mode (black), ELM-free NT (red) and PT L-mode (blue) discharges.}
    \label{fig:profcompare}
\end{figure}

For context, NT profiles are compared to traditional PT L-mode and NT H-mode profiles at matched plasma conditions in figure~\ref{fig:profcompare}. The PT H-mode reference case shown in black represents a typical Type-I ELMy H-mode on DIII-D (taken from \cite{nelson_time-dependent_2021}) and exhibits a strong pressure gradient in the edge characteristic of a typical H-mode pedestal. At matched $I_\mathrm{p}=1\,$MA, $B_\mathrm{t}=-2\,$T, $\langle n \rangle\sim5\times10^{20}\,$m$^{-3}$, $P_\mathrm{aux}\sim2.5\,$MW and $\beta_\mathrm{N}=1.7$, the red profile is characteristic of a typical ELM-free NT discharge. A small pressure pedestal exists in the NT plasma, though it is dwarfed by the large PB-unstable H-mode profile. Notably, the NT plasma supports steeper pressure gradients across the entire radius than the PT L-mode case shown in blue, which is from a discharge with matched $I_\mathrm{p}$, $B_\mathrm{t}$ and $P_\mathrm{aux}$, though at lower $\beta_\mathrm{N}$. As was observed in the profiles shown in figure~\ref{fig:profs-d}, the pressure gradient just inside of the NT pedestal is unusually steep (this is typical for NT plasmas on DIII-D) and remains steep all the way into the core plasma (though some flattening of the pressure gradient at $\psi_\mathrm{N}\lesssim0.4$ is also typical). Because of this, the ELM-free NT and PT H-mode cases compared in figure~\ref{fig:profcompare} attain similar volume-averaged pressures of $\langle p \rangle\sim23\,$kPa.


\subsection{Comparison with the EPED Model}
\label{subsec:EPED}

To better characterize the structure of the NT pedestal, a set of stationary phases is selected from the larger DIII-D dataset shown above for further analysis \cite{paz-soldan_simultaneous_2024}. These phases are automatically selected by identifying time windows of at least $\Delta t\gtrsim \mathrm{min(400\,ms, 3}\tau_\mathrm{e})$, where $\tau_\mathrm{e}$ is the energy confinement time, during which $\beta_\mathrm{N}$, $W_\mathrm{MHD}$ $q_\mathrm{95}$ and $\langle n \rangle$ are all \textit{quasi-stationary}. Here we define the quasi-stationary condition as requiring a normalized standard deviation $\sigma_\mathrm{\overline{x}}<3\%$ and a normalized average gradient $(dx/dt)\times(\Delta t/\overline{x}<0.25\%$, where $\overline{x}$ is the average of the quantity in question over the selected window. In addition to these constraints, stationary periods are omitted from the final dataset used in this work if they feature $\delta<-0.4$, $f_\mathrm{GW}>1$, are not diverted, are from experiments using impurity injection or studying internal transport barriers or feature significant $n=1$ or $n=2$ core activity. These selection criteria result in the identification of 328 unique periods covering the majority of the NT parameter space accessed in the DIII-D Campaign shape. These stationary periods are used for pedestal analysis throughout this section.

For H-mode discharges, the EPED1 model can be used to predict the pedestal height and width based upon the intersection of the non-local PB stability boundary with the nearly-local kinetic ballooning mode (KBM) constraint \cite{Snyder2009a, Snyder2011}. This model is often found to be in good ($\lesssim15-20\%$) agreement with experimental data and leads to a scaling of the form $\Delta_\mathrm{\psi_\mathrm{N}}=G\beta_\mathrm{\theta,ped}^{1/2}$, where $\Delta_\mathrm{\psi_\mathrm{N}}$ is the pedestal width in $\psi_\mathrm{N}$ space, $\beta_\mathrm{\theta,ped}$ is the pedestal poloidal beta and $G$ is a weakly varying function of normalized parameters with a typical value of $G=0.076$ for DIII-D \cite{Snyder2012}. In MKS units, $\beta_\mathrm{\theta,ped}$ is defined as $\beta_\mathrm{\theta,ped}\equiv2\mu_\mathrm{0}p_\mathrm{ped}/B_\mathrm{\theta}^2$, where $p_\mathrm{ped}$ is the pedestal top pressure and $B_\mathrm{\theta}$ is the flux-surface-averaged poloidal magnetic field at the separatrix. 

\begin{figure}
    \includegraphics[width=1\linewidth]{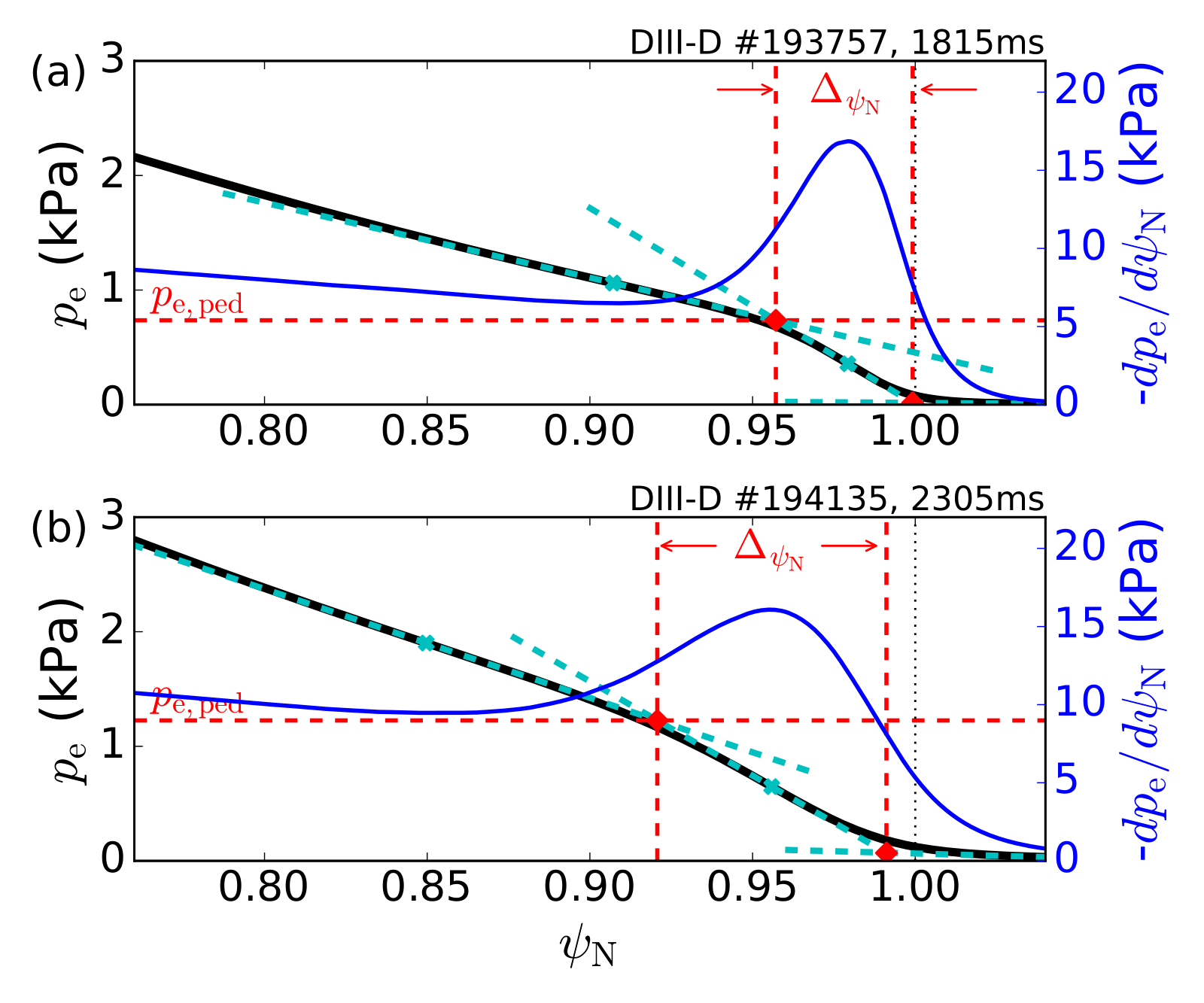}
    \labelphantom{fig:pedstruct-a}
    \labelphantom{fig:pedstruct-b}
    \caption{Examples of measured electron pressure (black) for (a) an NT discharge with a small steep pedestal and (b) an NT discharge with a broader, flatter profile. The pressure gradient is shown in blue, and tangent lines to the $p_\mathrm{e}$ profile at the $dp_\mathrm{e}/d\psi_\mathrm{N}$ peak and troughs (marked with X's) are shown in cyan. The pedestal width and height are identified at intersection points of the tangent lines and are marked in red.}
    \label{fig:pedstruct}
\end{figure}

To compare experimental results to the EPED1 model, it is necessary to parameterize the edge pressure profile in a systematic manner. For H-mode discharges, this is done by fitting the pressure profile with a modified tanh (\texttt{mtanh} \cite{Groebner2001}) function, such that the pedestal height can be identified at the asymptotic value of the tanh function inboard of the steep gradient region and the pedestal with as the difference between the inflection points in the tanh fit \cite{groebner_progress_2009}. While this approach is capable of adequately fitting the NT $p_\mathrm{e}$ profile in select cases, many of the NT $p_\mathrm{e}$ profiles on DIII-D follow a smoother trajectory for which the \texttt{mtanh} is not entirely suitable. In some instances, this smoother trajectory can still lead to higher edge pressures than cases with a distinct pedestal region, as illustrated in figures~\ref{fig:pedstruct-a} and \ref{fig:pedstruct-b}, which show the fitted $p_\mathrm{e}$ profile for NT cases with a small steep pedestal and a broader, flatter profile, respectively. Since all NT of the stationary NT cases identified for this analysis do feature some sort of gradient enhancement in the edge, an alternative scheme was developed to identify the pedestal height and width for NT discharge on DIII-D. This scheme is illustrated pictorially in figure~\ref{fig:pedstruct}. Tangent lines (cyan) are fit to the pressure profile at locations of the peak and troughs in the $dp_\mathrm{e}/d\psi_\mathrm{N}$ profile to capture the trajectory of the $p_\mathrm{e}$ profile (black). The pedestal width and height (red) are then determined by the intersection of these tangent curves. This results in a robust method for the determination of the $p_\mathrm{e}$ pedestal height and width for NT discharges that typically deviates by less than $\sim5\%$ from the conventional definition (\cite{groebner_progress_2009}) on H-mode discharges on DIII-D.

\begin{figure}
    \includegraphics[width=1\linewidth]{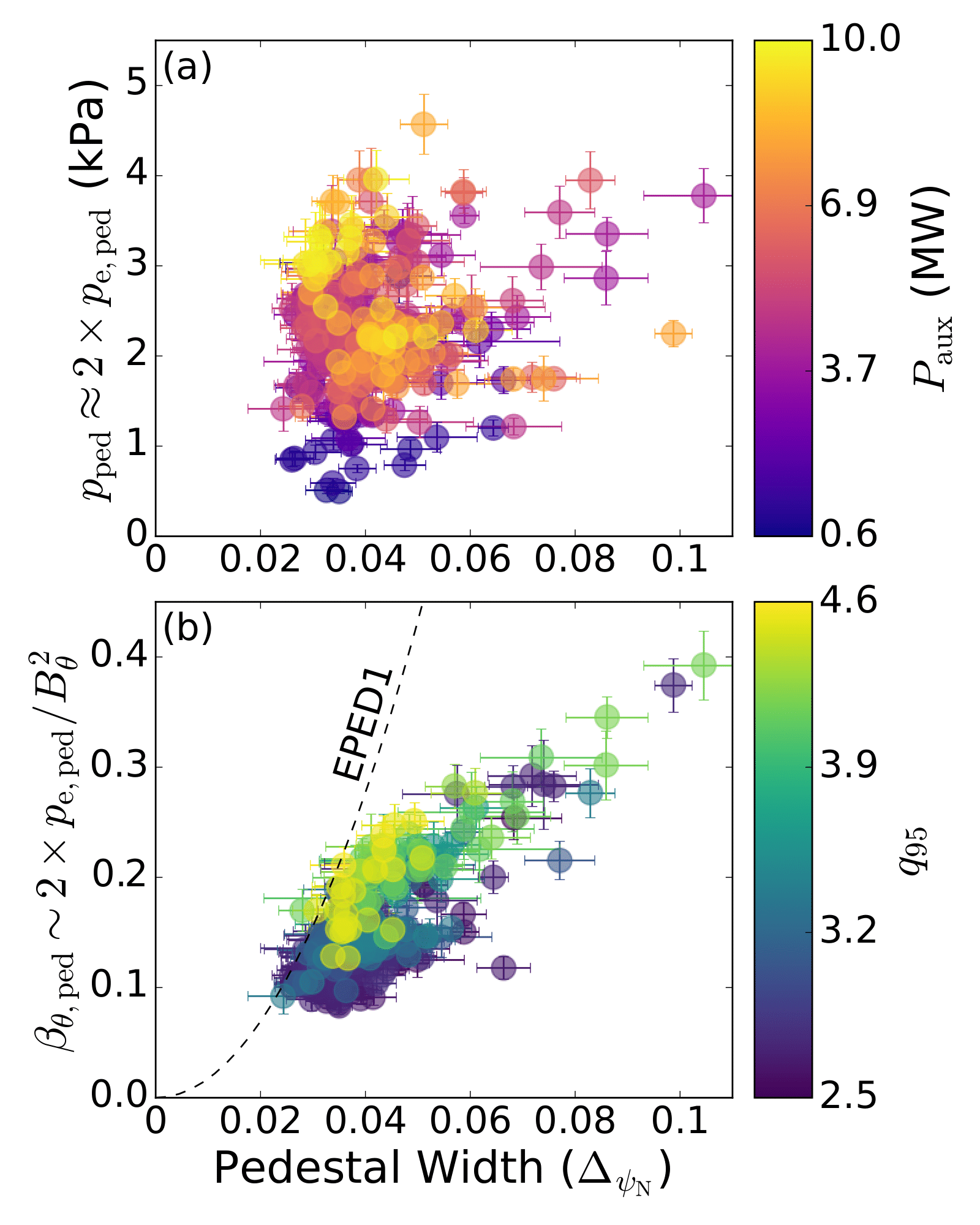}
    \labelphantom{fig:eped-a}
    \labelphantom{fig:eped-b}
    \caption{(a) Pedestal height (b) $\beta_\mathrm{\theta,ped}$ vs. the pedestal width $\Delta_\mathrm{\psi_\mathrm{N}}$ for the stationary phases of DIII-D NT discharges.}
    \label{fig:eped}
\end{figure}

Using this pedestal identification, in figure~\ref{fig:eped} we show the experimental electron pressure pedestal widths and heights for the stationary DIII-D dataset in comparison to the EPED1 scaling. Notably, the NT plasmas exhibit a significantly reduced $\beta_\mathrm{\theta,ped}$ (using $2\times p_\mathrm{e}$ for the pressure) compared to the EPED1 scaling for DIII-D, as expected from figure~\ref{fig:profcompare}. While $\beta_\mathrm{\theta,ped}$ is typically assumed to scale $\beta_\mathrm{\theta,ped}\sim\Delta_\mathrm{\psi_\mathrm{N}}^2$ in PT discharges, the relationship appears to be almost linear across the stationary NT dataset considered here with a different slope realized for each value of $q_\mathrm{95}$. This suggests that the physics models included in the EPED1 model (KBM and PB constraints) may not accurately describe the edge region in ELM-free NT plasmas. Notably, the most narrow NT pedestals ($\Delta_\mathrm{\psi_\mathrm{N}}\lesssim0.05$) tend to keep up with the EPED1 scaling, as could be inferred from the near-separatrix profiles presented in figure~\ref{fig:profcompare}, which are effectively matched for the NT and PT H-mode cases. In the NT dataset, this narrow region is typically characterized by substantial $E\times B$ shear which will be the subject of future publication.
However, in these cases the steep gradient is often not sustained further into the plasma and the PB constraint is never met, suggesting that agreement with EPED1 in this narrow region may be coincidental only. 

Previous work studying the stability of the NT edge has led to the development of the EPED-CH code, which uses a simple analytic expression for the KBM constraint as in EPED1 but does not include any diamagnetic stabilization for the PB modes and instead employs an iterative scheme to find the PB constraint through the execution of the CHEASE, CAXE and KINX codes \cite{merle_pedestal_2017}. Simulations with EPED-CH predict that the pedestal height for reactor-relevant NT discharges may be on the order of $\sim4$ times lower than `equivalent' PT discharges \cite{merle_pedestal_2017}. While they occupy a different parameter space than these predictions, the initial observations from DIII-D included in figure~\ref{fig:eped} suggest a somewhat smaller reduction (by a factor of $\sim2-3$) in the NT pedestal height compared to the EPED1 prediction. The EPED-CH model was not available for modeling of the currently presented discharges but will be included in future work on the subject. 

\subsection{Edge Effect on Core Performance}
\label{subsec:power}

A key benefit of NT plasmas is that they can sustain an inherently ELM-free edge while accessing high core performance. Several recent studies have identified gradient enhancement in the NT edge as a dominant effects leading to improved core confinement. For example, assessments of NT data from the TCV tokamak report that a pressure pedestal of $\rho>0.8$ is necessary to explain the significant confinement improvement observed with NT \cite{Sauter2014}. This helps to explain observations that, since the radial penetration depth of triangularity is finite, local gyrokinetic simulations only considering flux tubes within the core plasma often do not reproduce the confinement improvement observed in experiment \cite{marinoni_effect_2009, Camenen2007, merlo_investigating_2015}. Investigations of TCV NT plasmas with the GENE code have also demonstrated that local simulations can lead to a strong overestimation of the heat fluxes when $\delta<0$ \cite{merlo_nonlocal_2021}. Similarly, simulations with GENE and ASTRA-TGLF for NT scenarios on the Divertor Test Tokamak predict that significant gradient enhancement outside of $\rho=0.9$ is necessary for NT plasmas to recover the confinement levels of equivalent PT H-mode plasmas \cite{mariani_first-principle_2024}.

\begin{figure}
    \includegraphics[width=1\linewidth]{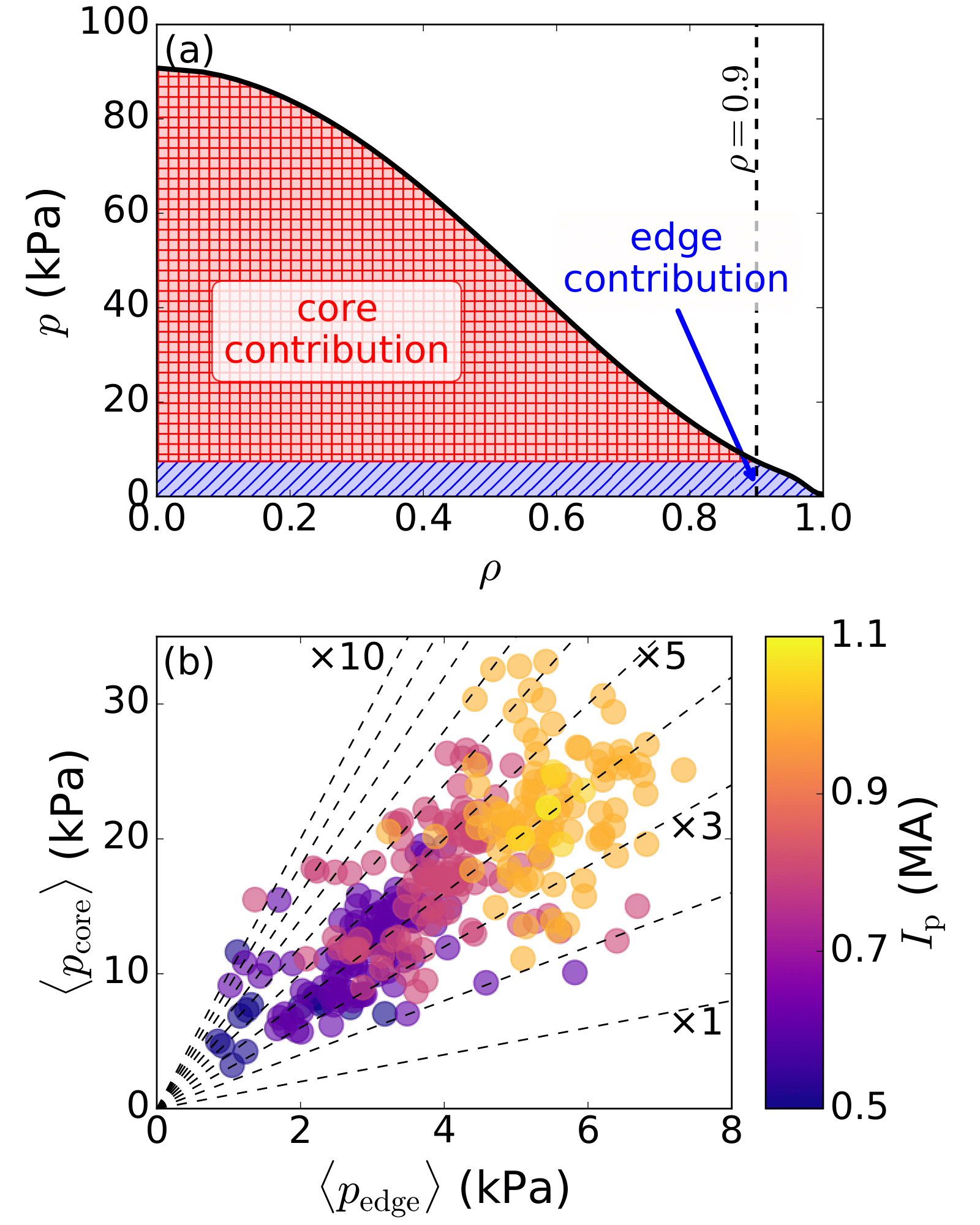}
    \labelphantom{fig:pdivide-a}
    \labelphantom{fig:pdivide-b}
    \caption{(a) An example pressure profile from an ELM-free NT discharge, divided into core and pedestal contributions. (b) Core and edge contributions to the volume-averaged pressure for a set of stationary ELM-free NT discharges on DIII-D, colored by $I_\mathrm{p}$. Constant factors $\langle p_\mathrm{core} \rangle / \langle p_\mathrm{edge} \rangle$ are shown as dashed lines.}
    \label{fig:pdivide}
\end{figure}

To begin to assess the importance of the NT edge in determining the performance of ELM-free NT plasmas on DIII-D, we plot the core and edge contributions to the volume-averaged pressure $\langle p \rangle$ for a set of stationary ELM-free NT discharges in figure~\ref{fig:pdivide}. Here $\langle p_\mathrm{core} \rangle$ and $\langle p_\mathrm{edge} \rangle$ are defined as the volume average of the pressure profile inside and outside of $\rho=0.9$, respectively, such that $\langle p_\mathrm{core} \rangle + \langle p_\mathrm{edge} \rangle \equiv \langle p \rangle$. We use the value $p(\psi_\mathrm{N}=0.9)$ as the edge pressure here to more consistently characterize the edge enhancement over a broad range of pedestal conditions seen in figure~\ref{fig:eped}, as steepened gradients often exist inside of the more narrow $p_\mathrm{e}$ pedestal characterized via the mechanisms described in figure~\ref{fig:pedstruct}. Across the DIII-D dataset, $\langle p_\mathrm{core} \rangle$ is strongly correlated with $\langle p_\mathrm{edge} \rangle$, even for discharges at the same $I_\mathrm{p}$. For the majority of cases, $\langle p_\mathrm{core} \rangle$ is $\sim4 - 5\times$ larger than $\langle p_\mathrm{edge} \rangle$. This sort of multiplicative scaling is expected from critical gradient theory \cite{Sauter2014}, suggesting that core transport is stiff in NT plasmas on DIII-D, consistent with degradation seen in scaling laws \cite{paz-soldan_simultaneous_2024}. As a result, NT plasmas benefit substantially from any enhancement of the edge gradient. 


\begin{figure}
    \includegraphics[width=1\linewidth]{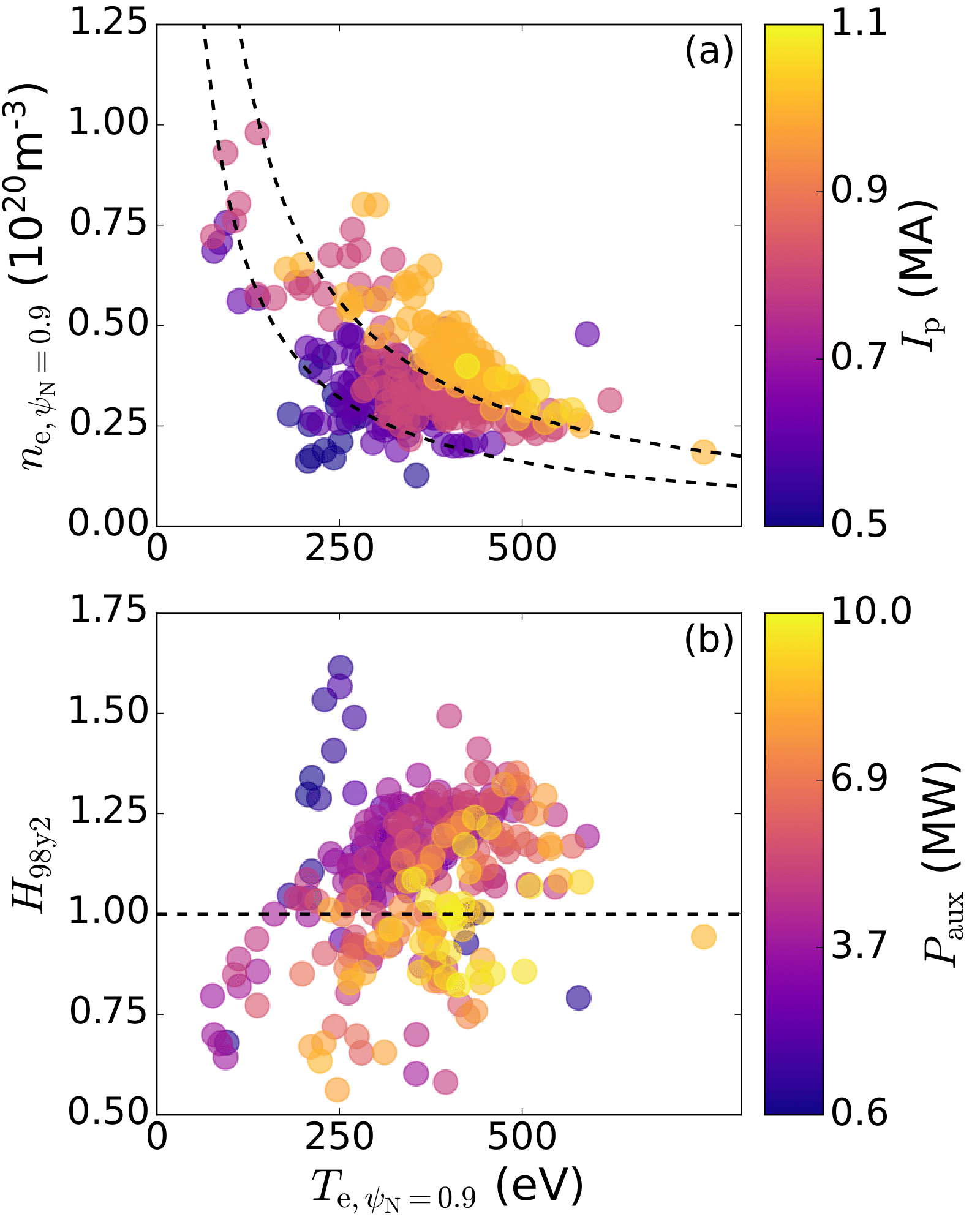}
    \labelphantom{fig:te95-a}
    \labelphantom{fig:te95-b}
    \caption{(a) Electron density and temperature at the $95\%$ flux surface, with lines of constant pressure marked. (b) Normalized confinement as a function of $T_\mathrm{e,95}$.}
    \label{fig:te95}
\end{figure}

The impact of the NT edge on plasma confinement is further examined in figure~\ref{fig:te95}, which shows the broad range of electron density and temperature values achievable in the NT edge as a function of plasma current. The maximum achievable edge pressure (marked by dashed curves in figure~\ref{fig:te95-a}) increases strongly with increasing $I_\mathrm{p}$, though select stationary cases with high pedestal pressures are also achieved at lower plasma currents. Notably, the primary lever for increasing edge pressure with $I_\mathrm{p}$ comes from the temperature channel, consistent with the strong $I_\mathrm{p}$ exponent in NT scaling laws \cite{paz-soldan_simultaneous_2024}. High edge densities are present in the DIII-D dataset at all $I_\mathrm{p}$ while lower current discharges are generally clustered at lower $T_\mathrm{e,\psi_\mathrm{N}=0.9}$. This suggests a strong scaling of edge temperature with $I_\mathrm{p}$, which is reminiscent of known scalings typical of PT H-mode discharges \cite{gruber_overview_1999, suttrop_identification_1997, Sauter2014}. Figure~\ref{fig:te95-b} shows a relatively strong correlation between $H_\mathrm{98y2}$ and $T_\mathrm{e,\psi_\mathrm{N}=0.9}$, similar to that previously reported in stationary enhanced $D_\mathrm{\alpha}$ (EDA) H-modes with no ELMs \cite{Hughes2011}. Across the dataset, increased $T_\mathrm{e,\psi_\mathrm{N}=0.9}$ leads to increased $H_\mathrm{98y2}$ at constant $P_\mathrm{aux}$ while increased $P_\mathrm{aux}$ leads to increased $T_\mathrm{e,\psi_\mathrm{N}=0.9}$ at constant $H_\mathrm{98y2}$. A general decrease in $H_\mathrm{98y2}$ is observed with increasing $P_\mathrm{aux}$ at constant $T_\mathrm{e,\psi_\mathrm{N}=0.9}$, indicative of core degradation at high heating powers. Notably, $T_\mathrm{e,\psi_\mathrm{N}=0.9}$ does not scale particularly strongly with $P_\mathrm{aux}$. The correlation between $H_\mathrm{98y2}$ and $T_\mathrm{e,\psi_\mathrm{N}=0.9}$ again emphasizes the extended benefit offered to NT plasma performance by sustaining edge profile gradients above traditional L-mode levels in a manner similar to ELMy and ELM-free H-mode discharges in PT. 

\section{Fluctuation Signatures in the NT Edge}
\label{sec:fluct}

The common discrepancy between the 1$^{st}$ stability limit and the normalized edge pressure gradient suggests the likely presence of additional gradient-limiting mechanisms in the NT edge on DIII-D. These could include high-to-moderate-$n$ interchange-like modes that still limit the plasma below the PB instability limit \cite{yu_understanding_2023}, electrostatic instabilities such as the Trapped Electron Mode (TEM) and Ion-Temperature Gradient Mode (ITG) or electromagnetic instabilities such as the Microtearing Mode (MTM) or Kinetic Ballooning Mode (KBM), among others. Further, DIII-D NT discharges of all shapes have been observed to host a variety of benign, non-disruptive tearing modes, among other classes of MHD instabilities, which could further play a role either in setting the maximum achievable $\beta_\mathrm{N}$ or in limiting pressure gradients throughout the plasma \cite{boyes_mhd_2023, cote_first_2024}. While work to identify and understand the turbulent mechanisms responsible for limiting the NT pedestal height and width is still ongoing, an initial assessment of common turbulence and transport signatures observed in DIII-D NT plasmas is presented in this section. 

\begin{figure}
    \includegraphics[width=1\linewidth]{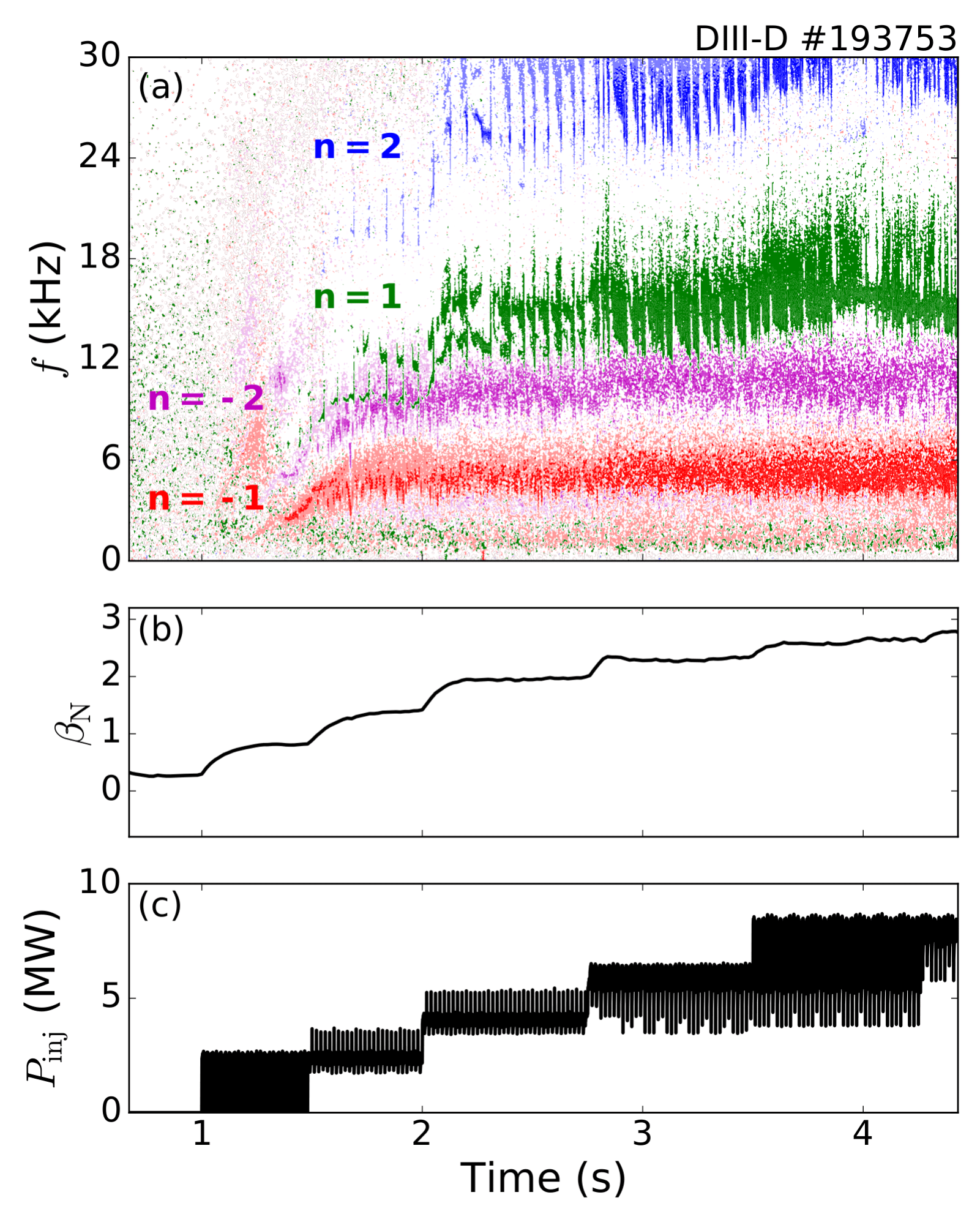}
    \labelphantom{fig:spec-a}
    \labelphantom{fig:spec-b}
    \labelphantom{fig:spec-c}
    \caption{(a) A typical spectrogram for a strongly-shaped NT discharge on DIII-D, showing two modes ($n=-1$ and $n=-2$) that are localized to the edge region. (b) $\beta_\mathrm{N}$ and (c) beam-injected power $P_\mathrm{inj}$ for this discharge increase over time.}
    \label{fig:spec}
\end{figure}

\begin{figure}
    \includegraphics[width=1\linewidth]{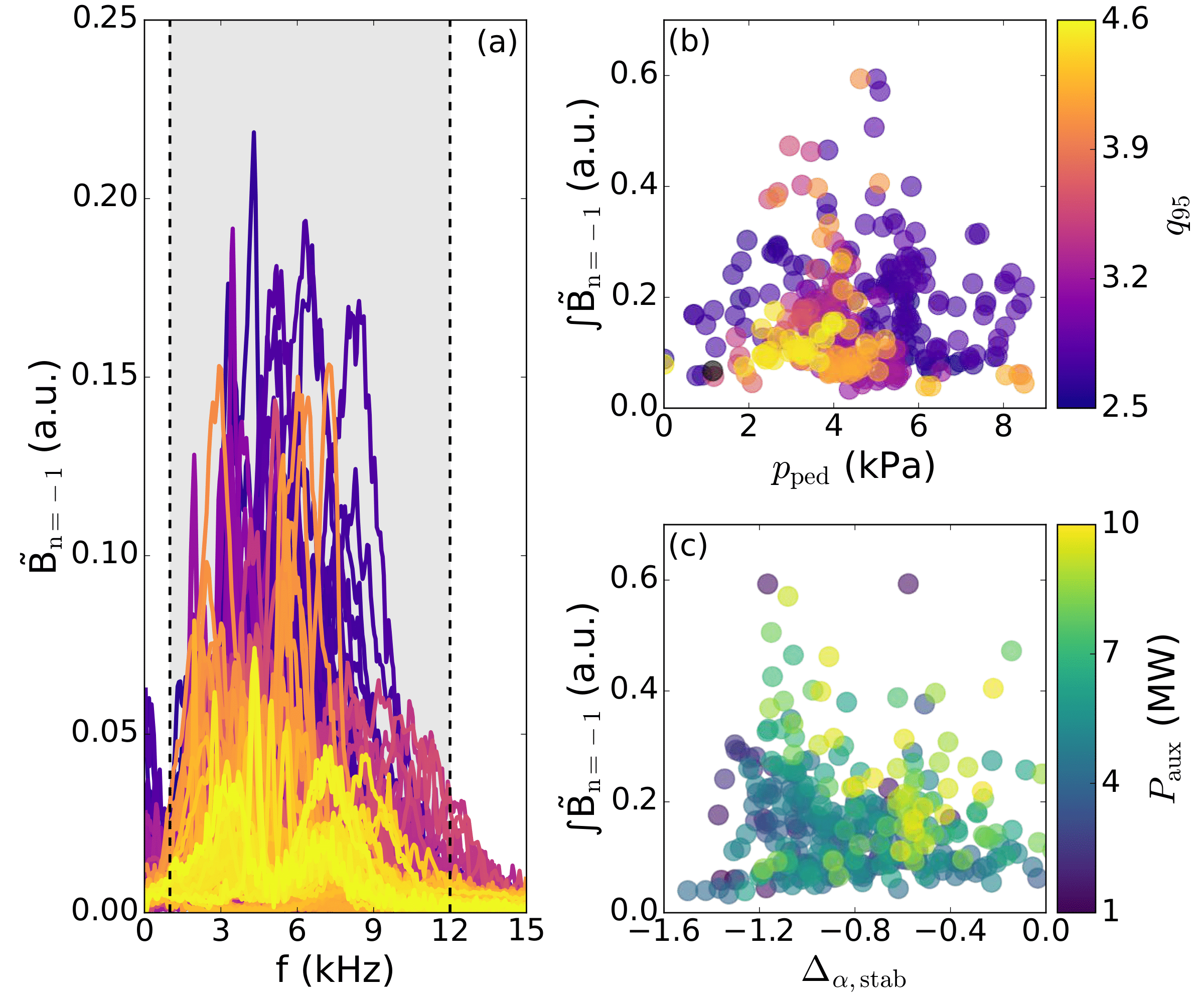}
    \labelphantom{fig:specint-a}
    \labelphantom{fig:specint-b}
    \labelphantom{fig:specint-c}
    \caption{(a) The time-averaged $n=-1$ fluctuation amplitude as a function of frequency for each stationary phase, colored by $q_\mathrm{95}$. The integrated $n=-1$ fluctuation amplitude -- using the gray shaded region in (a) as integration bounds -- does not correlate strongly with either (b) the pedestal pressure or (c) the distance to the ballooning stability boundary.}
    \label{fig:specint}
\end{figure}

In almost all of the campaign shapes run on DIII-D, an $n=-1$ edge mode appears in fast magnetic measurements. This is illustrated in figure~\ref{fig:spec} for a typical NT discharge, showing that the $n=-1$ mode appears as a distinct fluctuation signature compared to $n=1$ and $n=2$ sawteeth behavior, which are localized further towards the plasma core \cite{cote_first_2024}. Interestingly, while the $n=-1$ mode appears only after a critical $\beta_\mathrm{N}$ is achieved, it persists at a constant frequency across the discharge flattop and likely does not significantly hamper plasma confinement. As this particular fluctuation signature is unique to high-power, strongly-shaped NT discharges on DIII-D, it is a candidate for involvement in setting the edge profile gradients. In figure~\ref{fig:specint}, we calculated the total $n=-1$ fluctuation amplitude ($\int \tilde{B}_\mathrm{n-1}$) by integrating the time-averaged $n=1$ fluctuation amplitude $\tilde{B}_\mathrm{n=-1}(f)$ from 1 to 12\,kHz, as indicated by the grey shaded region in figure~\ref{fig:specint-a}. Importantly, no strong correlation is observed between $\int \tilde{B}_\mathrm{n-1}$ and either the pedestal height ($p_\mathrm{ped}$ -- figure~\ref{fig:specint-b}) or the distance to the 1$^\mathrm{st}$ stability boundary ($\Delta_\mathrm{\alpha,stab}$ -- figure~\ref{fig:specint-c}). This suggests that the $n=-1$ edge mode, while a consistent feature of NT discharges on DIII-D is likely not a signature of ballooning stability or of a pedestal-limiting effect. Further details describing and characterizing this mode can be found in \cite{cote_first_2024}.

\begin{figure}
    \includegraphics[width=1\linewidth]{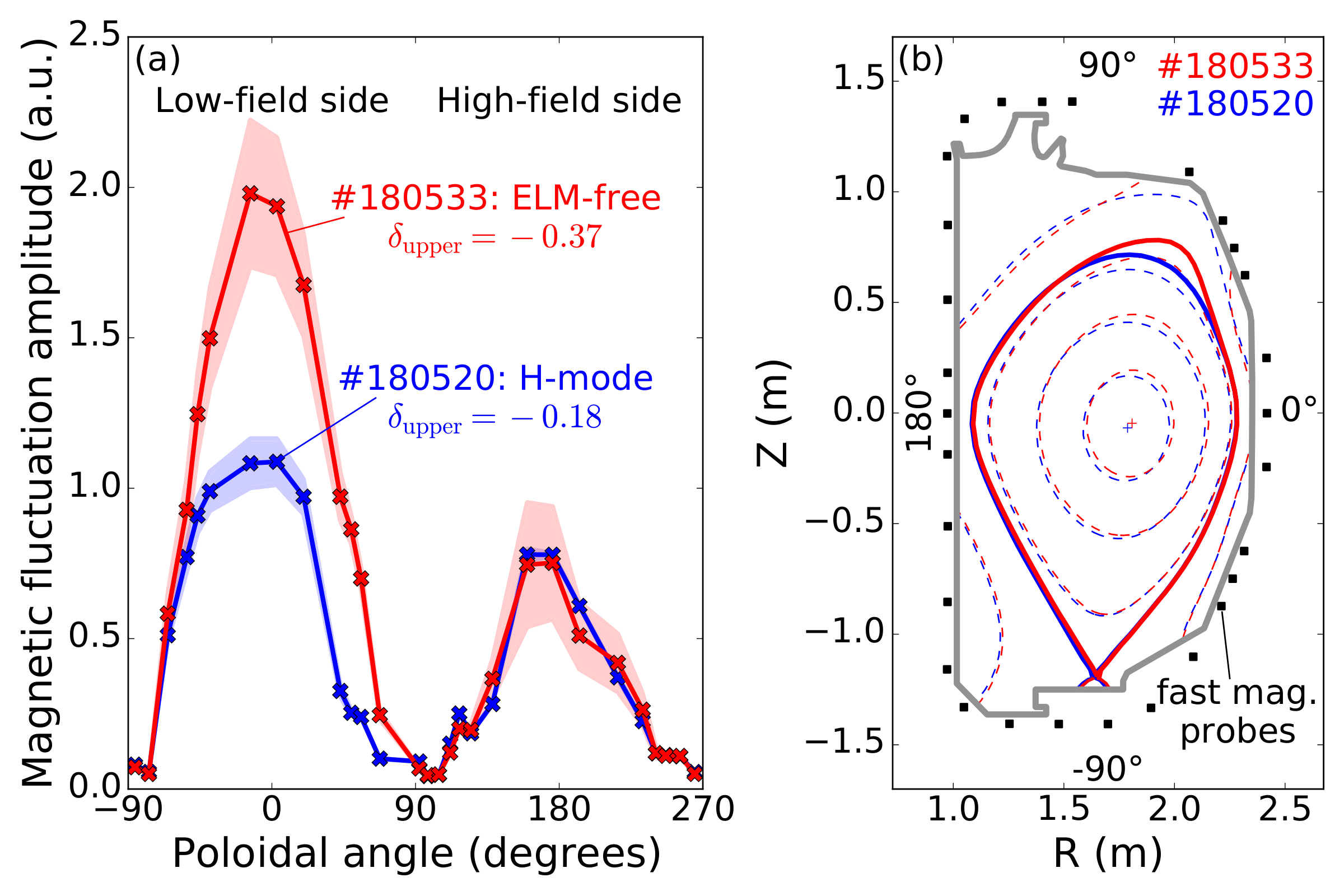}
    \labelphantom{fig:fluct-a}
    \labelphantom{fig:fluct-b}
    \caption{(a) Comparison of magnetic field fluctuation amplitude for two discharges at matched conditions: an ELMy H-mode discharge with $\delta_\mathrm{uppper}=-0.18$ (blue) and an ELM-free, ballooning-limited discharge with $\delta_\mathrm{upper}=-0.37$. (b) Equilibrium shapes and magnetic probe locations for these discharges are provided.}
    \label{fig:fluct}
\end{figure}

Another method with which to assess magnetic fluctuations prominent in DIII-D NT discharges is to compare the amplitude of broadband magnetic fluctuations between the low-field-side (LFS) and high-field-side (HFS). Since infinite-$n$ ballooning modes should be localized to the LFS, this manner of comparison can help to identify ballooning signatures. For two cases which straddle the critical $\delta$ needed for ELM-free operation at the same $I_\mathrm{p}$ in the reduced NT shape, the RMS amplitude of fast magnetic signals are plotted as a function of poloidal angle in figure~\ref{fig:fluct}. For reference, the two discharge shapes and the locations of each probe used in this analysis are provided in figure~\ref{fig:fluct-b}. Strikingly, the ELM-free discharge features significantly ($\sim\times2$) larger LFS  fluctuations, despite being matched in everything except $\delta$ to the ELMy H-mode case. While not conclusive evidence of the destabilization of infinite-$n$ ballooning modes at more negative $\delta$, this observation consistent with the ballooning theory. For clarity, we note that the suppressed fluctuation amplitudes observed at $\theta=\pm90^{\circ}$ are a result of the large separation between the plasma separatrix and the magnetic pickup coils at those locations, as seen in in figure~\ref{fig:fluct-b}. Since the outer midplane ($\theta=0^{\circ}$) separatrix location remains unchanged between these two discharges, the impact of probe distance on fluctuation amplitude at $\theta=0^{\circ}$ is negligible. 

Further studies of fluctuation signatures in other channels are ongoing (see references \cite{cote_first_2024} and \cite{stewart_characterization_2024}) and will not be reported in detail here, though a discussion of their impact on limiting gradient growth in the NT edge will be included in future work. Early studies examining DIII-D NT discharges with $\delta\sim\delta_\mathrm{crit}$ using electron cyclotron emission imaging diagnostics have identified low-$n$ pressure-driven resistive interchange-like modes in the plasma edge \cite{yu_understanding_2023}, though these have yet to be tied explicitly to gradient limits in the NT edge. Low-k density fluctuations studied using Beam Emission Spectroscopy \cite{mckee_beam_1999} indicate that drift-wave like electron-diamagnetic-directed modes dominate outside of $\rho\sim0.85$, consistent with trapped electron modes \cite{stewart_characterization_2024}. Their amplitudes peak around $\rho\sim0.95$ as they are locally suppressed at the separatrix, an effect often observed in H-modes. These modes are characterized by poloidally elongated correlation lengths ($L_{c \theta}> L_{cr}$) at the edge and broad wavenumber spectra. In scans of triangularity, low phase velocity, edge modes below 70$\,$kHz are observed in the low triangularity, H-mode phases; however, these modes are suppressed in the stronger triangularity, ELM-free phases, as is discussed in more depth in \cite{stewart_characterization_2024}. As discussed above, it is likely that while these additional instabilities or others may ultimately be responsible for clamping the edge pressure gradient under various specific conditions, leading to the spread of ballooning stability observed at all powers in figure~\ref{fig:balldb}, the high-$n$ ballooning mode remains as a eventual upper bound on the potential growth of the NT pedestal, ensuring that H-mode remains suppressed regardless of the plasma conditions.

\section{Conclusion}
\label{sec:conc}

All DIII-D NT discharges that have a \textit{sufficiently negative} triangularity (defined as $\delta<\delta_\mathrm{crit}\sim-0.15$, though the precise value of $\delta_\mathrm{crit}$ depends on the specifics of the discharge) are observed to be completely ELM-free, regardless of heating power (up to $P_\mathrm{aux}\gtrsim 8\times P_\mathrm{LH}$) or other machine parameters. Furthermore, it is sufficient for only one of the two triangularities ($\delta_\mathrm{top}$ and $\delta_\mathrm{bottom}$) to be past this critical threshold in order to robustly prevent ELMs. This allows for numerous cases within the DIII-D dataset to achieve an ELM-free state with average triangularities $\delta\sim0$, provided either that $\delta_\mathrm{top}<\delta_\mathrm{crit}$ or that $\delta_\mathrm{bottom}<\delta_\mathrm{crit}$. Since ELM-free operation in NT is a direct function of plasma shape, it is possible to achieve a wide range of scenarios while maintaining the ELM-free NT edge. These include scenarios with impressive normalized performance metrics \cite{paz-soldan_simultaneous_2024} and with unseeded divertor detachment \cite{scotti_high_2024}, among others. Optimization of the plasma shape to achieve the best core performance while retaining ELM-free operation will be a subject of future research on the DIII-D machine. 

The onset of the ELM-free NT state on DIII-D correlates directly with closure of the 2$^{nd}$ stability region for infinite-$n$ ballooning modes, suggesting that the ideal ballooning stability plays a critical role in preventing H-mode access and ELM events. Across the DIII-D dataset, all ELM-free NT plasmas feature normalized pressure gradients below the 1$^{st}$ stability limit for infinite-$n$ ballooning modes. It is important to note, however, that the pressure pedestal in the majority of NT discharges on DIII-D is not itself directly congruent with the 1$^{st}$ stability limit for infinite-$n$ ballooning modes; often the experimentally measured gradient sits somewhere below that which would be predicted unstable to the ideal ballooning mode. This is indicative of the presence of additional gradient-limiting mechanisms in the NT edge that prevents growth of the pressure pedestal and ultimately controls the pressure pedestal width and height in NT scenarios. Notably, the highest power cases are also not at this limit, suggesting that they may be degraded by some other mechanism. 

A host of fluctuations have been identified in the edge region of NT plasmas on DIII-D that may be signatures of these gradient-limiting instabilities. However, direct links between these signatures and the gradient-limiting mechanisms in the NT edge have yet to be conclusively identified. ELM-free NT plasmas on DIII-D are typically characterized by enhanced LFS magnetic fluctuations, consistent with (though not indicative of) the destabilization of high-$n$ ballooning modes on the outboard side. 
Further work to identify the dominant transport mechanisms responsible for setting the structure of the NT pedestal is needed. 

Despite this, the NT pedestal is well described across the DIII-D dataset through profile measurements. DIII-D NT plasmas are typically characterized by a steep, narrow temperature pedestal while the density channel often remains flatter, instead relying on an increase in separatrix density to achieve high values. A narrow shear layer (on the order of $\Delta_\mathrm{\psi_\mathrm{N}}\lesssim 2-5\%$) is typically present in the DIII-D NT edge, though wider `pedestals' reaching as far in as $\psi_\mathrm{N}\sim0.9$ are also observed. Due to the large variation in measured pedestal structures, we use density and temperature values at $\psi_\mathrm{N}=0.9$ as a standard metric for the `edge' conditions. With these values, it is found that the integrated core pressure scales almost linearly with the integrated edge pressure, such that $\langle p_\mathrm{core} \rangle \sim 4-5\times \langle p_\mathrm{edge} \rangle$ across the DIII-D dataset. This suggests that core transport in NT discharges remains relatively stiff such that the NT edge itself is responsible for a sizable contribution to the increase in core performance observed in NT plasmas, consistent with previous modeling results. 


The physics understanding of the NT edge is still evolving rapidly, and many open questions remain. For example, the establishment of a predictive scaling for the NT pedestal width and height based on first-principle models could be extremely lucrative for NT FPP design. Additionally, the experimental realization of optimized NT scenarios, which feature high-performance ELM-free plasmas in conjunction with a  detached edge, will go a long ways towards establishing the credibility of the NT regime as a reactor-relevant scenario. Efforts along these lines will be included in future work on the subject.

\section*{Acknowledgments}

The authors would like to particularly thank the large set of scientists that made the DIII-D NT campaign possible, as indicated in the author list of reference \cite{thome_overview_2024}. Some calculations performed in this study were completed through the OMFIT framework \cite{Meneghini2015}, and the generated data are available upon request. This material is based upon work supported by the U.S. Department of Energy, Office of Science, Office of Fusion Energy Sciences, using the DIII-D National Fusion Facility, a DOE Office of Science user facility, under Awards DE-SC0022270, DE-SC0022272, DE-SC0020287, DE-FC02-04ER54698, DE-AC02-09CH11466, DE-FG02-08ER54999, DE-FG02-97ER54415, DE-AC52-07NA27344, DE-AC05-00OR22725 and DE-SC0014264.

\section*{Disclaimer}

This report was prepared as an account of work sponsored by an agency of the United States Government. Neither the United States Government nor any agency thereof, nor any of their employees, makes any warranty, express or implied, or assumes any legal liability or responsibility for the accuracy, completeness, or usefulness of any information, apparatus, product, or process disclosed, or represents that its use would not infringe privately owned rights. Reference herein to any specific commercial product, process, or service by trade name, trademark, manufacturer, or otherwise does not necessarily constitute or imply its endorsement, recommendation, or favoring by the United States Government or any agency thereof. The views and opinions of authors expressed herein do not necessarily state or reflect those of the United States Government or any agency thereof.


\end{document}